\documentclass[aps,prc,showpacs,preprint,nofootinbib]{revtex4}

\usepackage{epsfig}
\usepackage{graphicx,amsmath}

\newcommand\ba{\begin{eqnarray}}
\newcommand\ea{\end{eqnarray}}
\newcommand\nn{\nonumber}
\newcommand{\be}{\begin{equation}}
\newcommand{\ee}{\end{equation}}

\begin{document}
\title{Polarization effects for the reaction $e^++e^-\to \bar p+p $ in presence of two--photon exchange - Lecture Notes}

\author{C. Adamu\v s\v c\'in}
\altaffiliation{Leave on absence from
\it Department of Theoretical Physics, IOP, Slovak Academy of Sciences, Bratislava, Slovakia}
\affiliation{\it DAPNIA/SPhN, CEA/Saclay, 91191 Gif-sur-Yvette Cedex, 
France}
\author{G. I. Gakh}
\altaffiliation{Permanent address:
\it NSC Kharkov Physical Technical Institute, 61108 Kharkov, Ukraine}
\affiliation{\it DAPNIA/SPhN, CEA/Saclay, 91191 Gif-sur-Yvette Cedex, 
France}
\author{E. Tomasi-Gustafsson}
\affiliation{\it DAPNIA/SPhN, CEA/Saclay, 91191 Gif-sur-Yvette Cedex, 
France}
\date{\today}
\begin{abstract}
The aim of this work is to give a pedagogical derivation of polarization observables for the annihilation reaction $e^+ + e^- \rightarrow N+ \bar N$. The reaction mechanism is one photon + two photon exchange, the last is described by axial parametrization. After deriving the general expressions for the cross section of a binary process, the matrix element is written in terms of three complex amplitudes. The method to derive polarization observables is detailed and all expressions are given in terms of generalized form factors. The strategy for determining physical form factors in annihilation reactions in presence of two photon exchange is suggested, on the basis of model independent properties of the relevant observables.
\end{abstract}

\maketitle
\tableofcontents

\section{Introduction}

The reactions $e ^{\pm} +p \to e^{\pm} +p$ and the crossed channel 
$e^++e^-\to p+\bar p$ and $p+\bar p\to e^++e^-$ are studied since many decades, as they are considered the simplest reactions which contain information on the nucleon structure.

Following the developments of the experimental possibilities: high intensity accelerators, polarized beams and targets, high resolution spectrometers, electron and hadron polarimeters, it has been only recently possible to measure polarization observables in space-like region, and to measure the annihilation cross section in a wide kinematical range in the time-like region. The physics goal is a precise extraction of hadron form factors in the full kinematical region. Evidently a comprehension of the reaction mechanism, including precise radiative corrections is necessary.

The aim of this work is to give a detailed description of a model independent formalism very well adapted to the extraction of cross section and polarization observables. The present results are focused on the reaction $e^++e^-\to p+\bar p$, as its experimental study is very actual.

Model independent expressions and statements are derived for the interesting experimental observables, when the reactions occur through the exchange of one and two photons.

\section{Differential cross section}
Let us define the cross section $\sigma $ for a binary process 
\be
a(p_1)+b(p_2) \to c(p_3)+d(p_4),
\label{eq:reac}
\ee
where the momenta of the particles are indicated in parenthesis. The cross section $\sigma$ characterizes the probability that a given process occurs. The number of final particles issued from a definite reaction is proportional to the number of incident particles $N_B$, the number of the target particles $ N_T$ and the constant of proportionality is the cross section:
\be
N_F=\sigma N_B \times N_T .
\label{defexpcrsec}
\ee

The cross section can be viewed as an 'effective area' over which the incident particle reacts. Therefore, its dimension is cm$^2$, but more often barn (1 barn=10$^{-28}$ m$^2$), or fm$^2$   (1 fm=10$^{-15}$ m). 

An useful quantity is the luminosity ${\cal L}$, defined as 
${\cal L}=N_B$ $[s^{-1}] N_T [cm^{-2}]$. For simple counting estimations, $N_f=\sigma {\cal L}$. This is an operative definition, which is used in experimental physics. 

On the other hand $\sigma$ needs to be calculated theoretically  for every type of process. The present derivation is done in a relativistic approach. This means that 
\begin{enumerate}
\item The kinematics is relativistic;
\item The matrix element ${\cal M}$, which contains the dynamics of the reaction is a relativistic invariant. In general it is function of kinematical variables, also relativistic ${\cal M}=f(s,t,u)$; 
\item $ \sigma$ has to be written in a relativistic invariant form;
\end{enumerate}
The starting point is the following expression for the cross section  
\be
d\sigma=\displaystyle\frac{ |{\cal M}|^2 }{\cal J} (2\pi)^4 \delta^{(4)}(p_1+p_2-p_3-p_4) d{\cal P},
\ee
which is composed by four terms:
\begin{enumerate}
\item The matrix element ${\cal M}$, which contains the dynamics of the reaction, and it is calculated following a model:
\item The flux of colliding particles ${\cal J}$;
\item The phase space for the final particles, $d{\cal P}$;
\item A term which insures the conservation of the four-momentum $\delta^{(4)}(p_1+p_2-p_3-p_4)$  which is the product of four 
$\delta$ functions, because each component has to be conserved separately.
\end{enumerate}
Let us calculate in detail each term.
\subsection{ Definition of flux}

The flux is defined through the relative velocity of incoming and target particles: 
\begin{subequations} \label{E:gp}
\begin{gather}
\mathcal{I}=n_B n_T v_{rel},   \label{E:gp1} \\
{\cal I}=4\sqrt{ (p_1\cdot p_2)^2-M_1^2M_2^2} ,  \label{E:gp2} 
\end{gather}
\end{subequations}
where $M_1$($M_2$) is the mass of the beam (target) particle, $v_{rel}$ is the relative velocity between beam and target particles and the densities of the beam and target particles $n_B, n_T$ are proportional to their energies as $n_i=2E_i$. 

Let us prove that the two expressions (\ref{E:gp1}) and (\ref{E:gp2}) are equivalent. It is more convenient to calculate $\mathcal{I}$ (Eq. \ref{E:gp} ) in the laboratory frame where the target is at rest:
\be
p_1=(E_1,\vec p_1), \; \; p_2=(M_2,0), \; \; |v_{rel}|=|\vec v_1-\vec v_2|= \frac{|\vec p_1|}{E_1} \; \Rightarrow \; n_B=2E_1, \; n_T=2M_2.
\label{eq:eq4}
\ee
Replacing the equalities (\ref{eq:eq4}) in  Eq. (\ref{E:gp1}):
$$\mathcal{I}=2E_1 2M_2\displaystyle\frac{|\vec p_1|}{E_1}=4M_2|\vec p_1|$$
and in Eq. (\ref{E:gp2}) :
$$(p_1\cdot p_2)^2-M_1^2M_2^2=M_2^2E_1^2-M_1^2M_2^2=M_2^2(E_1^2-M_1^2)=
M_2^2|\Vec p_1|^2, ~\to \mathcal{I}=4M_2 |\vec p_1|
$$
and the equalities (\ref{E:gp}) are proved. Moreover, we prove also that the flux does not depend on the reference frame, because it can be written in a Lorentz invariant form.

Let us consider the center of mass system (CMS): 
$$
p_1=(E_1,\vec k), \; p_2=(E_2,-\vec k),~
p_1\cdot p_2=E_1 E_2 + |\vec k|^2, \; M_1^2=E_1^2-|\vec k|^2, \; M_2^2=E_2^2-|\vec k|^2
$$ 
and
\ba
(p_1\cdot p_2)^2-M_1^2 M_2^2&=&E_1^2 E_2^2 + 2E_1 E_2 |\vec k|^2+|\vec k|^4-E_1^2 E_2^2 + |\vec k|^2 (E_1^2+E_2^2)-|\vec k|^4\nn \\
&=&|\vec k|^2(E_1+E_2)^2=|\vec k|^2 W^2.
\ea
The flux, $\mathcal{I}$, can be written as 
\be
\mathcal{I}=4 |\vec k|W, \label{fluxf}
\ee
where $W=E_1+E_2$ is the initial energy of the system in CMS.

\subsection{ Phase space}

The phase space for a particle of energy $E$, mass $M$ and four--momentum $p$ (the number of states in the unit volume) can be written from quantum mechanics in an invariant form:
$$d{\cal P}= \int \displaystyle\frac{d^4 p~\delta (p^2-M^2)}{(2\pi)^3} \Theta(E)$$
where the $\delta $ function insures that the particle is on mass shell 
and the step function $\Theta(E)$ insures that only the solution with positive energy is taken into account. Note that the wave functions of all particles entering in the matrix element must be normalized to one particle per unit volume. In this case all these wave functions contain the factor $1/\sqrt{2\varepsilon}$, where $\varepsilon $ is the particle energy.
Usually these factors are explicitly taken into account in the expression
for the cross section, we insert them into the phase space.

Extracting the term which depends on energy:
$$d^4p~\delta (p^2-M^2)=\delta^3\vec p dE \delta (E^2-\vec p^2-M^2).
$$
and using the property of the $\delta$ function
\be
\int \delta[f(x)]dx=\sum\displaystyle\frac{1}{|f'(x_i)|},
\label{eq:delta}
\ee
($x_i$ are the roots of $f(x)$),  with $f(E)=E^2-\vec p^2-M^2$, and $f'(E)=2E$ one finds:
$$\int dE\delta (E^2-\vec p^2-M^2)\Theta(E)=\displaystyle\frac{1}{2E}$$
For the considered reaction: 
$$d{\cal P}=\displaystyle\frac{d^3\vec p_3}{(2\pi)^3 2E_3}
\displaystyle\frac{d^3\vec p_4}{(2\pi)^3 2E_4}.$$

\subsection{Calculation of the cross section}
The total cross section can be written as:
\be
\sigma =  \frac{(2\pi)^4}{{\cal I}}
\int |\mathcal{M}|^2 \delta^{(4)}(p_1+p_2-p_3-p_4) \frac{d^3 \vec p_3}{(2\pi)^3 2E_3}\frac{d^3 \vec p_4}{(2\pi)^3 2E_4}. \label{defthcrsec}
\ee
One can see that it corresponds to a six-fold differential, but four $\delta$ functions are equivalent to four integrations. So finally, for a binary process one is left with two independent variables, $(E,\theta)$ or $(s,t)$. For three particles, one has nine differentials, four integrations, i.e., five independent variables.

The term $\delta^{(4)}(p_1+p_2-p_3-p_4)$  can be split into an energy and a space part:
$\delta^{(4)}(p_1+p_2-p_3-p_4)=\delta(E_1+E_2-E_3-E_4)
\delta^{(3)}(\vec p_1+\vec p_2-\vec p_3-\vec p_4)$. 

Note that
\be
	\int \delta^{(3)} (\vec p_1+\vec p_2-\vec p_3-\vec p_4) d^3 \vec p_4 = 1
\ee
in any reference frame.

Let us use spherical coordinates in  CMS ($p_3=(E_3,\vec p)$, $p_4=(E_4,-\vec p)$, $d^3\vec p=|\vec p|^2 d \Omega d p$)and consider the quantity ${\cal J}$:
\be
	{\cal J}=\delta(E_1+E_2-E_3-E_4)\frac{d^3 \vec p_3}{4E_3 E_4} = \delta (W-E_3 - E_4) \frac{|\vec p|^2 d \Omega d p}{4E_3 E_4}, \label{medzik}
\ee
where 
$$E_3^2=M_3^2+|\vec p|^2, \; E_4^2=M_4^2+|\vec p|^2 \to E_3 dE_3 = E_4 dE_4 = |\vec p| dp.$$
After integration, using the property (\ref{eq:delta}):
\be
	{\cal J}=\int \delta (W-E_3 - E_4)\displaystyle\frac{d E_3|\vec p| 
	d\Omega }{4E_4}= \displaystyle\frac{|\vec p| d\Omega}{4E_4} \displaystyle\frac{1}
	{\left|\displaystyle\frac{d}{d E_3}
	\left (W-E_3-E_4\right )\right|},
\ee
where 
\be
	\displaystyle\frac{d}{d E_3}(W-E_3-E_4)=-1-\displaystyle\frac{d E_4}{d E_3}=-1-\frac{E_3}{E_4}=-\displaystyle\frac{W}{E_4}
\ee
and therefore
\be
	{\cal J}=\frac{|\vec p| d\Omega}{4W}. \label{intak}
\ee

Substituting Eqs. (\ref{fluxf}, \ref{intak}) in  Eq. (\ref{defthcrsec}) we find the general expression for the differential cross section of a binary process, in CMS:
\be
\frac{d\sigma}{d\Omega}= \frac{|\mathcal{M}|^2|\vec p|}{64 \pi^2 W^2 |\vec k|},
\ee
and for the total cross section:
\be
	\sigma=\int \frac{|\mathcal{M}|^2|\vec p|}{64 \pi^2 W^2 |\vec k|}d\Omega.
\ee
In case of elastic scattering, $|\vec k|=|\vec p|$, therefore:
\be
\frac{d\sigma}{d\Omega}^{el}= \frac{|\mathcal{M}|^2}{64 \pi^2 W^2 }=|{\cal F}^{el}|^2
\ee
with the elastic amplitude  ${\cal F}^{el}=\displaystyle\frac{|\mathcal{M}|}{8 \pi W }$. 

For the annihilation reaction considered here, $e^+ +e^- \to N +\bar N$, neglecting the  mass of the electron, one has: 
$$|\vec k|=\frac{W}{2}, \; 
|\vec p|=\sqrt{E^2-M^2}=E\sqrt{1-M^2/E^2}=\frac{W}{2}\beta,$$
and 
\be
	\frac{d\sigma}{d\Omega}^{ann}= 
	\frac{|\mathcal{M}|^2 \beta}{64\pi^2 q^2},
	\label{finalcrossec}
\ee
where $\beta=\sqrt{1-4M^2/q^2}$ and 
$ q^2=s= (p_1+ p_2)^2$.

\section{Axial parametrization of the matrix element}
\begin{figure}
	\centering
		\includegraphics{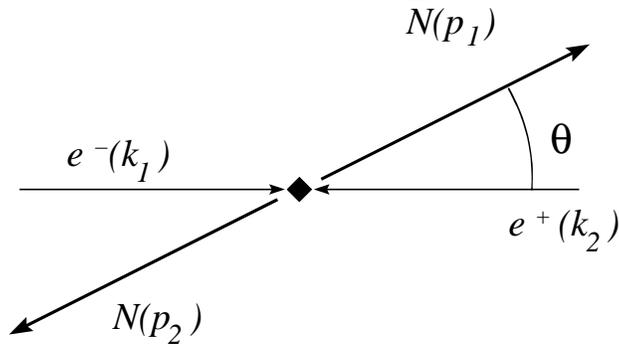}
	\caption{\small{Annihilation $e^- + e^+ \rightarrow \bar N + N$ in CMS system.}}
	\label{fig:eeNNcms}
\end{figure}

In presence of two photon exchange (TPE), the matrix element of the reaction $e^{-}(k_1)+e^{+}(k_2)\to \bar N(p_1)+N(p_2)$, can be parametrized by three complex amplitudes. In the present derivation we will use the following expression for the matrix element of this reaction, taking into account the TPE  contribution, 
\ba
\mathcal{M} &=&-\frac{e^2}{q^2}\bigg\{ \bar{u}(-k_2)\gamma_{\mu}u(k_1)\bar{u}(p_2)\left[ F_1(q^2,t)\gamma_{\mu}-\frac{F_2(q^2,t)}{2m} \sigma_{\mu\nu} q_{\nu} \right] u(-p_1) \nn \\
&&+ \bar{u}(-k_2)\gamma_{\mu} \gamma_5 u(k_1)\bar{u}(p_2) \gamma_{\mu} \gamma_5 u(-p_1)A_{2\gamma}(q^2,t) \bigg\},\label{eq:mat} 
\ea
where $m$ is nucleon mass, $k_1$ and $k_2$ are electron and positron four-momenta, $p_1$ and $p_2$ are antinucleon and nucleon four-momenta, $q$ is the four momentum of the virtual photon and $t$ is the Mandelstam variable, the momentum transfer: $q=k_1+k_2=p_1+p_2$ and $t=(k_1-p_1)^2$. The first two amplitudes contain the contributions of $1\gamma\bigotimes 2\gamma$ exchange, whereas the third amplitude is fully induced by $2\gamma$ exchange. $A_{2\gamma}(q^2,t)$ can be parametrized in different but equivalent ways. Here we use the axial parametrization that  describes the exchange of a $1^+$ particle. The spin and parity of the transition induced by TPE can be any, but the C-parity must be positive (whereas it is negative for $1\gamma$ exchange). 

The three complex amplitudes, $F_{1,2}(q^2, t)$ and $A_{2\gamma}(q^2, t),$ which 
generally are functions of two independent kinematical variables, $q^2$ 
and $t$, fully describe the spin structure of the matrix element 
for the reaction $ e^++e^-\rightarrow N+\bar N$  - for any number of exchanged virtual photons, because they contain C-odd and C-even terms.

This expression (\ref{eq:mat}) holds under assumption of the P--invariance of the
electromagnetic interaction and conservation of lepton helicity, which is 
correct for standard QED at the high energy, i.e., in zero electron mass limit. 
Note, however, that expression (\ref{eq:mat}) is one of the many equivalent representations
of the  $e^++e^-\rightarrow N+\bar N$ reaction matrix element.

In the Born ($1\gamma$ exchange) approximation these amplitudes reduce to:
\be\label{eq:eq3}
 F_1^{Born}(q^2, t)=F_1(q^2), \ 
 F_2^{Born}(q^2, t)=F_2(q^2), \
 A_{2\gamma}^{Born}(q^2, t)=0, \ 
\ee
where $F_1(q^2)$ and $F_2(q^2)$ are the Dirac and Pauli nucleon
electromagnetic form factors (FFs), respectively, and they are complex functions of the variable $q^2$. The complexity of FFs arises from the 
final--state strong interaction of the produced $N\bar N-$pair. In the 
following we use the standard magnetic $G_M(q^2)$ and charge $G_E(q^2)$ 
nucleon FFs which are related to FFs $F_1(q^2)$ and $F_2(q^2)$ as follows
\be\label{eq:eq4a}
 G_M=F_1+F_2, \  G_E=F_1+\tau F_2, \  
 \tau =\frac{q^2}{4m^2}>0. 
\ee
By analogy with these relations, let us introduce a linear combinations of the
$F_{1,2}(q^2, t)$ amplitudes which in the Born approximation correspond to the Sachs FFs $G_M$ and $G_E$:
\ba
\tilde G_M(q^2, t)&=&F_1(q^2, t)+F_2(q^2, t), \nn\\
\tilde G_E(q^2, t)&=&F_1(q^2, t)+\tau F_2(q^2, t). 
\label{eq:eq5}
\ea
The matrix element (\ref{eq:mat}) can be rewritten in terms of vector and axial electromagnetic currents:
\be
\mathcal{M}=-\frac{e^2}{q^2}\left(j_{\mu}^{(v)}J_{\mu}^{(v)}+j_{\mu}^{(a)}J_{\mu}^{(a)}\right),\label{matel2}
\ee
where $j_{\mu}^{(v)}$, $j_{\mu}^{(a)}$ are vector and axial lepton currents and $J_{\mu}^{(v)}$, $J_{\mu}^{(a)}$ are vector and axial nucleon currents:
\ba
&j_{\mu}^{(v)}=\bar{u}(-k_2)\gamma_{\mu}u(k_1), 
&J_{\mu}^{(v)}=\bar{u}(p_2)\left[ F_1(q^2,t)\gamma_{\mu}-\frac{F_2(q^2,t)}{2m} \sigma_{\mu\nu} q_{\nu} \right] u(-p_1), \nn \\
&j_{\mu}^{(a)}=\bar{u}(-k_2)\gamma_{\mu} \gamma_5 u(k_1),
&J_{\mu}^{(a)}=\bar{u}(p_2) \gamma_{\mu} \gamma_5 u(-p_1)A_{2\gamma}(q^2,t), 
~\sigma_{\mu\nu}=\frac{1}{2}[\gamma_{\mu},\gamma_{\nu}].
\label{matcur}
\ea
Then the differential cross section of the reaction $e^{-}+e^{+}\rightarrow \bar N+N$ in CMS according to (\ref{finalcrossec}) can be written as
\begin{eqnarray*}
\frac{d\sigma}{d\Omega}&=&\frac{\alpha^2\beta}{4q^6}\left(j_{\mu}^{(v)}J_{\mu}^{(v)}+j_{\mu}^{(a)}J_{\mu}^{(a)}\right)\left(j_{\nu}^{(v)}J_{\nu}^{(v)}+j_{\nu}^{(a)}J_{\nu}^{(a)}\right)^*\\
&=&\frac{\alpha^2\beta}{4q^6}\left[L_{\mu\nu}^{(v)}H_{\mu\nu}^{(v)}+2Re(L_{\mu\nu}^{(i)}H_{\mu\nu}^{(i)})\right],~ \alpha=\frac{e^2}{4\pi}=\frac{1}{137},
\end{eqnarray*}
where we neglected terms proportional to $A_{2\gamma}^2$ (since the amplitude $A_{2\gamma}$ is entirely due to the TPE contribution, which is of the order of $\alpha$). The  'vector' $(v)$ and 'interference' $(i)$ leptonic/hadronic tensors are defined as
\be
L_{\mu\nu}^{(v)}=j_{\mu}^{(v)}j_{\nu}^{(v)*}, \: L_{\mu\nu}^{(i)}=j_{\mu}^{(a)}j_{\nu}^{(v)*}, \: H_{\mu\nu}^{(v)}=J_{\mu}^{(v)}J_{\nu}^{(v)*}, \: H_{\mu\nu}^{(i)}=J_{\mu}^{(a)}J_{\nu}^{(v)*}. \label{matelform}
\ee
Note that the term proportional to the Dirac FF, $F_1$, in the expression for the nucleon vector current, $J_{\mu}^{(v)}$, (\ref{matcur}), is gauge invariant, when both particles ($N,\bar N$) are on mass shell. The second term proportional to the Pauli FF, $F_2$, is always gauge invariant:
$$(\sigma_{\mu\nu}q_{\nu})q_{\mu}=\frac{1}{2}(\gamma_{\mu}\gamma_{\nu} - \gamma_{\nu}\gamma_{\mu})q_{\nu}q_{\mu}=\frac{1}{2}(\hat q \hat q -\hat q \hat q)=0.$$
It is possible to find other forms of the nucleon vector current $J_{\mu}^{(v)}$, which are equivalent only for on-shell particles. In our case nucleons are the final particles, therefore they are on-shell.

Let us show that for on-shell nucleons the expression for the $J_{\mu}^{(v)}$ (\ref{matcur}) can be simplified by using Dirac equations\footnote{It is correct only when nucleon and antinucleon are on mass shell (real particles), i.e., they satisfy the Dirac equation.} for particles (nucleon - $p_2$) and antiparticles (antinucleon - $p_1$)
\ba
\bar{u}(p_2)(\hat{p}_2-m)=0 &\Rightarrow& \bar{u}(p_2) \hat{p}_2 = \bar{u}(p_2)m \nn \\
(\hat{p}_1+m)u(-p_1)=0 &\Rightarrow& \hat{p}_1u(-p_1)=-u(-p_1)m \nn
\ea
and the properties of Dirac matrices :
$\{\gamma_{\mu},\gamma_{\nu}\}=2g_{\mu\nu} $,
where $g_{\mu\nu}$ is the metric tensor of the Minkowski spacetime,
$\hat a\hat b+\hat b\hat a=2ab$, $ \hat a\gamma_{\mu}+ \gamma_{\mu}\hat a=2a_{\mu}$, where $a$ and $b$ are four vectors.

Let us develop the term accompanying $F_2$:
\ba
\bar{u}(p_2)\sigma_{\mu\nu}q_{\nu}u(-p_1)&=&\displaystyle\frac{1}{2}
\bar{u}(p_2)\left (\gamma_{\mu}\gamma_{\nu}-\gamma_{\nu}\gamma_{\mu}\right )q_{\nu}
u(-p_1)
=\displaystyle\frac{1}{2}\bar{u}(p_2)
\left (\gamma_{\mu} \hat q-\hat q\gamma_{\mu}\right ) u(-p_1)\nn \\
&=&\displaystyle\frac{1}{2}
\bar{u}(p_2)\left [\gamma_{\mu}(\hat p_1+\hat p_2)-(\hat p_1+\hat p_2)\gamma_{\mu}\right ]
u(-p_1)
\nn \\
&=&
\displaystyle\frac{1}{2}
\bar{u}(p_2)\left [\gamma_{\mu} (-m+\hat p_2)-(\hat p_1+m)\gamma_{\mu}\right ]
u(-p_1)\nn \\
&=&\displaystyle\frac{1}{2}\bar{u}(p_2)\left [-2m\gamma_{\mu}+( \gamma_{\mu}\hat p_2 -\hat p_1\gamma_{\mu})\right ]u(-p_1)\nn \\
&=&\displaystyle\frac{1}{2}\bar{u}(p_2)\left [-2m\gamma_{\mu}+(2p_{2\mu}-\hat p_2\gamma_{\mu}-2p_{1\mu}+\gamma_{\mu}\hat p_1)\right] u(-p_1)
\nn\\
&=&\displaystyle\frac{1}{2}\bar{u}(p_2)\left [-4m\gamma_{\mu}+2(p_2-p_1)_{\mu}\right] u(-p_1).
\ea
Replacing in the expression for $J_{\mu}^{(v)}$, Eq. (\ref{matcur}):
\be 
J_{\mu}^{(v)} =\bar{u}(p_2)\left[ (F_1+F_2) \gamma_{\mu} - \frac{F_2}{2m}(p_2-p_1)_{\mu}
\right] u(-p_1)=\bar{u}(p_2)\left[ \left( F_1+F_2 \right) \gamma_{\mu} - \frac{F_2}{m} P_{\mu} \right] u(-p_1),
\label{deriv}
\ee
where $P=(p_2-p_1)/2$ and $F_1$, $F_2$ can be substituted by generalized magnetic and charge nucleon FFs, Eq. (\ref{eq:eq5}):
\be
J_{\mu}^{(v)}=\bar{u}(p_2)\left[ \tilde G _M(q^2,t) \gamma_{\mu} - \frac{\tilde G_M(q^2,t)-\tilde G_E(q^2,t)}{m(1-\tau)} P_{\mu} \right] u(-p_1). \label{hadronV}
\ee 
For simplicity, we will use in our calculations:
\be
\frac{\tilde G_M -\tilde G_E }{m(1-\tau)}=G_2.
\label{eq:eqfft}
\ee

\section{Lepton and Hadron Tensors}
We give a detailed derivation of the tensors, in particular of the lepton tensor and of the matrix components. 

\subsection{Lepton tensors}
The calculation of the leptonic tensors leads to the calculation of a trace. Let us  give the explicit derivation. From Eqs. (\ref{matcur},\ref{matelform}), the expression for the 'vector part' of the leptonic tensor is:
\be
L_{\mu\nu}^{(v)}=\bar{u}(-k_2) \gamma_{\mu} u(k_1) \left[ \bar{u}(-k_2) \gamma_{\nu} u(k_1) \right]^{*} \label{Lmunuvector}.
\ee
Using the definition $\bar u(-k_2)=u^\dagger(-k_2)\gamma_4=u^*(-k_2)\gamma_4$ and the following properties of the  $\gamma$ matrices: $\gamma_4^*=\gamma_4$,
$(\gamma_4)_{ij}=(\gamma_4)_{ji}$, $(\gamma_4)_{kl}(\gamma_4)_{lm}=\delta_{km}$, the complex conjugated term can be written as
\be
\left[ \bar{u}(-k_2) \gamma_{\nu} u(k_1) \right]^{*}=\left[ u^*(-k_2)\gamma_4 \gamma_{\nu} u(k_1) \right]^{*}=u(-k_2)\gamma_4^* \gamma_{\nu}^* u^*(k_1).
\ee
In component form (with spinor indices): 
\ba
u_i(-k_2)(\gamma_4^*)_{ij} (\gamma_{\nu}^*)_{jk} u^*(k_1)_k
&=&u(-k_2)_i(\gamma_4)_{ij} (\gamma_{\nu}^*)_{jk} \delta_{km}u^*(k_1)_m
\nn\\
&=&u(-k_2)_i(\gamma_4)_{ij}
 (\gamma_{\nu}^*)_{jk}(\gamma_4)_{kl}(\gamma_4)_{lm}u^*(k_1)_m\nn\\
&=&
u^*_m(k_1)(\gamma_4)_{ml}(\gamma_4)_{lk}(\gamma_{\nu}^{\dagger})_{kj} (\gamma_4)_{ji}
u_i(-k_2)\nn\\
&=&\bar u(k_1) \gamma_4 \gamma_{\nu}^{\dagger} \gamma_4 u(-k_2)
=\bar u(k_1) \gamma_{\nu} u(-k_2).\nn
\ea
Therefore
\be
\left[ \bar{u}(-k_2) \gamma_{\nu} u(k_1) \right]^{*}=\bar u(k_1) \gamma_{\nu} u(-k_2).
\ee
This result will be used all along the paper, with other terms between bispinors ($\gamma_{\nu}, \gamma_{\nu} \gamma_5, P_{\nu}$).

Let us write the tensor (\ref{Lmunuvector}) in component form 
\ba
L_{\mu\nu}^{(v)}&=&\bar{u}_i(-k_2) (\gamma_{\mu})_{ij} u(k_1)_j \bar u_a(k_1) (\gamma_{\nu})_{ab} u_b (-k_2)
=u_b(-k_2) \bar{u}_i(-k_2) (\gamma_{\mu})_{ij} u_j(k_1) \bar u_a(k_1) (\gamma_{\nu})_{ab} \nn \\
&= &(\rho_2)_{bi} (\gamma_{\mu})_{ij}  (\rho_1)_{ja} ( \gamma_{\nu})_{ab} 
=Tr [ u(-k_2) \bar{u}(-k_2) \gamma_{\mu} u(k_1) \bar u(k_1) \gamma_{\nu}],\label{Lmunuvectorfinal}
\ea
where we applied the property that a product of matrices is a matrix and the first and last indices coincide: $Tr \mbox{A}=\sum_b \mbox{A}_{bb}$. The density matrices $\rho=u(p)\bar u(p)$ for polarized and unpolarized particles and antiparticles are given in the Table \ref{tab:Density}. 

\begin{table}[htbp]
	\centering
		\begin{tabular}{|c|c|c|}
		\hline
		&particle&antiparticle\\
		\hline
		unpolarized&$ \hat p +m $&$ \hat p -m$\\
		\hline
		polarized&$ (\hat p +m)\frac{1}{2}(1-\gamma_5 \hat s)$&$(\hat p -m)\frac{1}{2}(1-\gamma_5 \hat s)$\\
		\hline
		\end{tabular}
	\caption{\small{The density matrices for polarized/unpolarized particles and antiparticles.}}
	\label{tab:Density}
\end{table}
  
The polarization four-vector $s$ is related to the unit vector along polarization of the particle in its  rest system, $\vec{\xi}$ by
\be
s_0=\frac{1}{m}\vec{p}\cdot \vec{\xi} \: ; \: \vec{s}=\vec{\xi} + \frac{\vec{p}(\vec{p}\cdot \vec{\xi})}{m(m+E)}. \label{4pol}
\ee

Let us consider firstly unpolarized incoming positron and longitudinally polarized incoming electron. In this case the leptonic vector tensor, can be written as
\be
L_{\mu\nu}^{(v)}=Tr\left[(\hat{k}_2-m_e) \gamma_{\mu} (\hat{k}_1+m_e) \frac{1}{2}(1-\gamma_5 \hat{s}) \gamma_{\nu}\right]=L_{\mu\nu}^{(v)}(0)+L_{\mu\nu}^{(v)}(S)
\ee
and expanded as a sum over polarization states.

\subsubsection{ The unpolarized lepton tensor : $L_{\mu\nu}^{(v)}(0)$}
Let us extract the part of the leptonic vector tensor which does not depend on polarization:
\be
L_{\mu\nu}^{(v)}(0)=\frac{1}{2}Tr\left[(\hat{k}_2-m_e) \gamma_{\mu} (\hat{k}_1+m_e) \gamma_{\nu} \right]=\frac{1}{2}\left[ Tr (\hat{k}_2 \gamma_{\mu} \hat{k}_1 \gamma_{\nu} ) - m_e^2Tr\left(\gamma_{\mu} \gamma_{\nu} \right) \right]. \nn
\ee
Using the rules for calculating the traces of Dirac matrices : 
$Tr \gamma_{\mu}\gamma_{\nu}=4g_{\mu\nu}$ and $Tr \gamma_{\rho} \gamma_{\mu} \gamma_{\sigma} \gamma_{\nu}=4(g_{ \rho\mu}g_{\sigma\nu}+
 g_{\mu\sigma}g_{\nu\rho}-g_{\sigma\rho}g_{ \mu\nu})$ one finds:
\be
L_{\mu\nu}^{(v)}(0)=2\left( k_{1\nu}k_{2\mu}+k_{1\mu}k_{2\nu}-k_1.k_2 g_{\mu\nu}-m_e^2 g_{\mu\nu}\right)=-q^2g_{\mu\nu} + 2\left(k_{1\nu}k_{2\mu}+k_{1\mu}k_{2\nu}\right), \label{Lvmunu0}
\ee
where we used the identity
\be
q^2=(k_1+k_2)^2=k_1^2+2k_1k_2+k_2^2=2(m_e^2+k_1k_2)\Rightarrow k_1k_2+m_e^2=\frac{q^2}{2}.
\ee
The tensor describing unpolarized electrons is symmetric.
\subsubsection{The polarized lepton tensor : $L_{\mu\nu}^{(v)}(S)$}

For the polarized part of the lepton tensor one has
\ba
L_{\mu\nu}^{(v)}(S)&=&-\frac{1}{2}Tr\left[(\hat{k}_2-m_e) \gamma_{\mu} (\hat{k}_1+m_e) \gamma_5 \hat{s} \gamma_{\nu}\right] 
=-\frac{1}{2}m_e\left\{Tr\left[\gamma_5 \hat{k}_2 \gamma_{\mu} \hat{s} \gamma_{\nu} \right] - Tr\left[\gamma_5 \gamma_{\mu} \hat{k}_1 \hat{s} \gamma_{\nu} \right]\right\} \nn \\
&=&
2m_e i \left\langle k_2 \mu s \nu \right\rangle - 2m_e i \left\langle\mu k_1 s \nu \right\rangle = 2m_e i \left\langle \mu \nu s q \right\rangle, \label{Lmini}
\ea
where we used the notation 
$$Tr~\gamma_5\gamma_{\mu}\gamma_{\nu}
\gamma_{\rho}\gamma_{\sigma}=-4i\varepsilon_{\mu\nu\rho\sigma} =-4i
\left\langle \mu\nu\rho\sigma \right\rangle, $$
and the properties of permutations of Dirac matrices. The Greek letters $\mu$, $\nu$,... are used for the non contracted indices of the antisymmetric tensor $\varepsilon_{\mu\nu\rho\sigma}$. 

One can check that the tensor $L_{\mu\nu}^{(v)}(S)$ (\ref{Lmini}) has the following property, which follows from current conservation:
$$q_{\mu}\cdot L_{\mu\nu}^{(v)}(S)=\varepsilon_{\mu\nu\sigma\rho}s_{\sigma}q_{\rho}q_{\mu}=0$$
as it is the product of an antisymmetric tensor ($\varepsilon_{\mu\nu\sigma\rho}$) times a symmetric tensor $q_{\rho}q_{\mu}$.

When the electron is longitudinally polarized ($\vec{\xi} \parallel \vec{k}_1 \to \vec{\xi}\cdot\vec{k}_1=|\vec{k}_1|=\sqrt{E^2-m_e^2} \approx E$), the components of the polarization vector $s_{\mu}$ (Eq. \ref{4pol}) become
\be
s_0=\frac{E}{m_e} \: ; \: \vec{s}=\vec{\xi}\left(1+\frac{|\vec{k}_1|^2}{m_e(m_e+E)}\right)=\vec{\xi}\left(1+\frac{E^2-m_e^2}{m_e(E+m_e)}\right)=\vec{\xi}\frac{E}{m_e}, \mbox{ i.e., } s_{\mu}=\lambda_e \frac{k_{1\mu}}{m_e}, \label{longpolel}
\ee 
where the helicity $\lambda_e$ takes the values $=\pm 1$ if $\vec{\xi}$ is parallel or antiparallel to $\vec k_1$. One can see that the longitudinally polarized part of 'vector' lepton tensor (\ref{Lmini}) is not suppressed by the  electron mass and it can be written as:
\be
L_{\mu\nu}^{(v)}(S)=2i \lambda_e \left\langle \mu \nu k_1 q \right\rangle. \label{Lvmini1}
\ee
Notice that the transversal component of the vector polarization remains unchanged and should be evaluated from (\ref{Lmini}).

\subsubsection{ The unpolarized lepton tensor : $L_{\mu\nu}^{(i)}(0)$}
According to (\ref{matcur}) and (\ref{matelform})
\be
L_{\mu\nu}^{(i)}=\bar{u}(-k_2)\gamma_{\mu} \gamma_5 u(k_1) \left[ \bar{u}(-k_2)\gamma_{\nu}u(k_1) \right]^*=\bar{u}(-k_2)\gamma_{\mu} \gamma_5 u(k_1)\bar{u}(k_1)\gamma_{\nu}u(-k_2), \nn
\ee
resp.
\ba
L_{\mu\nu}^{(i)}&=&Tr\left[u(-k_2)\bar{u}(-k_2)\gamma_{\mu} \gamma_5 u(k_1)\bar{u}(k_1)\gamma_{\nu}\right] \nn \\
&=&Tr\left[(\hat{k}_2-m_e)\gamma_{\mu} \gamma_5 (\hat{k}_1+m_e)\frac{1}{2}(1-\gamma_5 \hat{s})\gamma_{\nu}\right].
\ea
Again it can be divided to polarized and unpolarized part. For the unpolarized part
\be
L_{\mu\nu}^{(i)}(0)=\frac{1}{2}Tr\left[(\hat{k}_2-m_e)\gamma_{\mu} \gamma_5 (\hat{k}_1+m_e)\gamma_{\nu}\right]=\frac{1}{2}Tr\left[\gamma_5 \hat{k}_2 \gamma_{\mu} \hat{k}_1 \gamma_{\nu}\right],
\ee
which can be expressed as
\be
L_{\mu\nu}^{(i)}(0)=\frac{1}{2}(-4i)\varepsilon_{\rho\mu\sigma\nu}k_{2\rho}k_{1\sigma}=2i\left\langle \mu \nu k_2 k_1 \right\rangle. \label{Limini0}
\ee

\subsubsection{The polarized lepton tensor : $L_{\mu\nu}^{(i)}(S)$}

The polarized part of $L_{\mu\nu}^{(i)}$ is written as: 
\ba
L_{\mu\nu}^{(i)}(S)&=&-\frac{1}{2}Tr\left[(\hat{k}_2-m_e)\gamma_{\mu} \gamma_5 (\hat{k}_1+m_e) \gamma_5 \hat{s} \gamma_{\nu}\right] \nn \\
&=&-\frac{1}{2}Tr\left[(\hat{k}_2-m_e)\gamma_{\mu} m_e \hat{s} \gamma_{\nu}\right] + \frac{1}{2}Tr\left[(\hat{k}_2-m_e)\gamma_{\mu}  \hat{k}_1 \hat{s} \gamma_{\nu}\right], \label{Lminiis}
\ea
where we used $\gamma_5^2=1$. Eq. (\ref{Lminiis}) can be simplified to
\begin{eqnarray*}
L_{\mu\nu}^{(i)}(S)&=&-\frac{m_e}{2}\left[ Tr(\hat{k}_2 \gamma_{\mu} \hat{s} \gamma_{\nu}) - Tr(\gamma_{\mu} \hat{k}_1 \hat{s} \gamma_{\nu}) \right] \\
&=& -2m_e \left[ k_{2\mu} s_{\nu} + k_{2\nu} s_{\mu} - k_2\cdot s g_{\mu\nu} -  k_{1\mu} s_{\nu} + k_{1\nu} s_{\mu} - k_1\cdot s g_{\mu\nu} \right].
\end{eqnarray*}
In case of longitudinally polarized electron beam, with the help of Eq. (\ref{longpolel}), this expression simplifies to:
\be
L_{\mu\nu}^{(i)}(S)=\lambda_e[q^2 g_{\mu\nu} - 2 (k_{2\mu} k_{1\nu} + k_{2\nu} k_{1\mu})]. \label{Limini1}
\ee

\subsubsection{Lepton tensor summary}

The leptonic tensors for the case of longitudinally polarized electrons
\ba
L_{\mu\nu}^{(v)}&=&-q^2 g_{\mu\nu} + 2 (k_{1\mu} k_{2\nu}+k_{1\nu}k_{2\mu})+2i\lambda_e\left\langle \mu \nu k_1 q \right\rangle \nn \\
L_{\mu\nu}^{(i)}&=&2i\left\langle \mu \nu k_2 k_1 \right\rangle + \lambda_e [ q^2 g_{\mu\nu} - 2 (k_{1\mu} k_{2\nu}+k_{1\nu}k_{2\mu}) ],
\ea
where $\lambda_e$ is the degree of the electron longitudinal polarization. We will consider that the lepton is fully polarized, i.e., $|\lambda_e|=1$, but it  shows explicitly which part of the leptonic tensor depends on polarization of the incoming electron.

\subsection{Hadron tensors}
According to the definitions (\ref{matelform}) and (\ref{hadronV}), $H_{\mu\nu}^{(v)}$ can be expressed as
\ba
H_{\mu\nu}^{(v)} &=& \bar{u}(p_2) \left[\tilde G_M \gamma_{\mu} - G_2 P_{\mu} \right] u(-p_1) \left[ \bar{u}(p_2) \left[\tilde G_M \gamma_{\nu} - G_2 P_{\nu} \right] u(-p_1) \right]^* \nn \\
&=& \bar{u}(p_2) \left[\tilde G_M \gamma_{\mu} - G_2 P_{\mu} \right] u(-p_1) \bar u(-p_1) \left[ \tilde G_M^* \gamma_{\nu} -G_2^* P_{\nu} \right] u(p_2) \label{Hminicele} \\
&=& Tr \left[ u(p_2) \bar{u}(p_2) \left[\tilde G_M \gamma_{\mu} - G_2 P_{\mu} \right] u(-p_1) \bar u(-p_1) \left[\tilde  G_M^* \gamma_{\nu} -G_2^* P_{\nu} \right] \right]. \nn
\ea
Generally, taking into account the polarization states of the produced nucleon and antinucleon, the hadronic tensor can be written as the sum of three contributions
\be
H_{\mu\nu}=H_{\mu\nu}(0)+H_{\mu\nu}(s_1)+H_{\mu\nu}(s_1,s_2),
\ee
where the tensor $H_{\mu\nu}(0)$ describes the production of unpolarized particles, the tensor $H_{\mu\nu}(s_1)$ describes the production of polarized nucleon or antinucleon and the tensor $H_{\mu\nu}(s_1,s_2)$ corresponds to the production of both polarized particles ($N$ and $\bar{N}$). 

According to this notation and with the help of the  expressions of the density matrices from Table \ref{tab:Density}, Eq. (\ref{Hminicele}) can be written as:
\be
H_{\mu\nu}^{(v)} = Tr \left \{ (\hat p_2 +m ) \left[ \tilde G_M \gamma_{\mu} - G_2 P_{\mu} \right] (\hat p_1 -m) \frac{1}{2} (1-\gamma_5 \hat s_1) \left[ \tilde G_M^* \gamma_{\nu} - G_2^* P_{\nu} \right] \right\}, \label{Hminiceleb}
\ee
which can be considered as a sum of polarized and unpolarized parts (similarly to the leptonic tensor), $s_{1\mu}$ is the polarization four-vector of the antinucleon.

\subsubsection{ The unpolarized hadron tensor : $H_{\mu\nu}^{(v)}(0)$}

The unpolarized part of $H_{\mu\nu}^{(v)}$ can be extracted from (\ref{Hminiceleb})
\begin{eqnarray*}
H_{\mu\nu}^{(v)}(0) &=& \frac {1}{2} Tr \left[ (\hat p_2 +m ) \left(\tilde G_M \gamma_{\mu} - G_2 P_{\mu} \right) (\hat p_1 -m) \left ( \tilde G_M^* \gamma_{\nu} - G_2^* P_{\nu} \right)  \right] \\
&=& \frac {1}{2} \Big[ \tilde G_M \tilde G_M^* Tr(\hat p_2 \gamma_{\mu} \hat p_1 \gamma_{\nu} ) + m \tilde G_M G_2^* P_{\nu} Tr ( \hat p_2 \gamma_{\mu} ) + G_2  G_2^* P_{\mu} P_{\nu} Tr( \hat p_2 \hat p_1) \\
&&+ m \tilde G_M^* G_2 P_{\mu} Tr ( \hat p_2 \gamma_{\nu} ) - m \tilde G_M G_2^* P_{\nu} Tr(\gamma_{\mu} \hat p_1) -m^2\tilde  G_M \tilde G_M^* Tr(\gamma_{\mu} \gamma_{\nu} )  \nn \\
&&- m \tilde G_M^* G_2 P_{\mu} Tr (\hat p_1 \gamma_{\nu} ) - m^2 G_2 G_2^* P_{\mu} P_{\nu} Tr \hat 1  \Big]
\end{eqnarray*}
where we omit the terms containing an odd number of $\gamma$ matrices, since their trace vanishes, and further simplify as:
\begin{eqnarray*}
H_{\mu\nu}^{(v)}(0) &=& 2 \Big[ |\tilde G_M|^2 (p_{1\mu}p_{2\nu}+p_{1\nu}p_{2\mu}-(p_1 p_2+m^2) g_{\mu\nu}) \\
&&+ P_{\mu} P_{\nu} \big (|G_2|^2 (p_1 p_2 -m^2) + 4m Re \tilde G_M G_2^* \big) \Big].
\end{eqnarray*}
Now we can apply following identities
\be
p_1 p_2 +m^2=\frac{q^2}{2} \: ; \: p_1 p_2 - m^2 = 2 m^2 (\tau -1) \: ; \: p_{1\mu}p_{2\nu}+p_{1\nu}p_{2\mu}=\frac{q_{\mu}q_{\nu}}{2}-2P_{\mu}P_{\nu} \label{idenhad}
\ee
to obtain
\be
H_{\mu\nu}^{(v)}(0) = H_1 \tilde{g}_{\mu\nu} + H_2 P_{\mu} P_{\nu}, \label{Hvmunu0}
\ee
where $\tilde{g}_{\mu\nu} = g_{\mu\nu} - q_{\mu} q_{\nu}/q^2$ and
\ba
H_1 &=& -q^2 |\tilde G_M|^2\\
H_2 &=& 4 \left [ m^2 |G_2|^2 (\tau - 1) - |\tilde G_M|^2 +2 m Re (\tilde G_M G_2^*) \right ]. \nn
\ea
The formula for $H_2$ can be rewritten in terms of $G_M$ and $G_E$
\be
H_2 = \frac{4}{\tau-1} \left [ |\tilde G_E|^2 - \tau |\tilde G_M|^2 \right ].
\ee

\subsubsection{The polarized hadron tensor : $H_{\mu\nu}^{(v)}(s_1)$}
The polarized part of $H_{\mu\nu}^{(v)}$ (also from (\ref{Hminiceleb}))
\ba
H_{\mu\nu}^{(v)}(s_1) &=& -\frac{1}{2} Tr \left[ (\hat p_2 +m ) \left(\tilde G_M \gamma_{\mu} - G_2 P_{\mu} \right) (\hat p_1 -m) \gamma_5 \hat s_1 \left( \tilde G_M^* \gamma_{\nu} -G_2^* P_{\nu} \right) \right] \nn \\
&=& \frac{1}{2} \Big[ Tr ( \hat p_2 \tilde G_M \gamma_{\mu} \hat p_1 \gamma_5 \hat s_1 G_2^* P_{\nu} ) + Tr ( \hat p_2 \tilde G_M \gamma_{\mu} m \gamma_5 \hat s_1  \tilde G_M^* \gamma_{\nu} ) \nn \\
&&+ Tr ( \hat p_2 G_2 P_{\mu} \hat p_1 \gamma_5 \hat s_1 \tilde G_M^* \gamma_{\nu} ) - Tr ( m \tilde G_M \gamma_{\mu} \hat p_1 \gamma_5 \hat s_1 \tilde G_M^* \gamma_{\nu} ) \nn \\
&=& \frac{1}{2} \Big[ - \tilde G_M G_2^* P_{\nu} Tr ( \gamma_5 \hat p_2 \gamma_{\mu} \hat p_1 \hat s_1 ) + m |\tilde G_M|^2 Tr ( \gamma_5 \hat p_2 \gamma_{\mu} \hat s_1 \gamma_{\nu}) \nn \\
&&+ \tilde G_M^* G_2 P_{\mu} Tr ( \gamma_5 \hat p_2 \hat p_1 \hat s_1 \gamma_{\nu} ) - m |\tilde G_M|^2 Tr ( \gamma_5 \gamma_{\mu} \hat p_1 \hat s_1 \gamma_{\nu}) \Big] \nn
\ea
can be simplified
\ba
H_{\mu\nu}^{(v)}(s_1) &=& 2i \Big[ \tilde G_M G_2^* P_{\nu} \left\langle p_2 \mu p_1 s_1 \right\rangle +m |\tilde G_M|^2 ( \left\langle \mu p_1 s_1 \nu \right\rangle - \left\langle p_2 \mu s_1 \nu \right\rangle ) \nn \\
&&- \tilde G_M^* G_2 P_{\mu} \left\langle p_2 p_1 s_1 \nu \right\rangle  \Big] \nn \\
&=& 2i \Big[ (\tilde G_M G_2^*)^* P_{\mu} \left< \nu p_2 p_1 s_1 \right> - \tilde G_M G_2^* P_{\nu} \left< \mu p_2 p_1 s_1 \right>\nn + m|\tilde G_M|^2 \left< \mu \nu q s_1 \right> \Big],\nn
\ea
or alternatively in terms of $\tilde G_M$ and $\tilde G_E$
\ba
H_{\mu\nu}^{(v)}(s_1) &=& \frac{2}{m(\tau-1)} \Big[ i m^2 (\tau - 1) |\tilde G_M|^2 \left< \mu \nu q s_1 \right> \nn \\
&&+ i Re(\tilde G_M (\tilde G_E-\tilde G_M)^*) ( P_{\mu} \left< \nu p_2 p_1 s_1 \right> - P_{\nu} \left< \mu p_2 p_1 s_1 \right> ) \nn \\
&&+ Im (\tilde G_M \tilde G_E^*) (P_{\mu} \left< \nu p_2 p_1 s_1 \right> + P_{\nu} \left< \mu p_2 p_1 s_1 \right> ) \Big], \label{Hvmini1}
\ea
where we used $Im |G_M|^2=0$ and identity
\ba
Re (A) (P_{\mu} Q_{\nu} -P_{\nu} Q_{\mu}) - i Im (A) ( P_{\mu} Q_{\nu} + P_{\nu} Q_{\mu} ) =A^* P_{\mu} Q_{\nu} - A P_{\nu} Q_{\mu}, \nn
\ea
which can be easily proved. Notice, that the first 2 terms in equation (\ref{Hvmini1}) are antisymmetric and the third (last) term is symmetric with respect to the exchange $\mu \leftrightarrow \nu$. 

\subsubsection{The unpolarized hadron tensor : $H_{\mu\nu}^{(i)}(0)$}

Using the definitions (\ref{matelform}) and (\ref{hadronV})
\ba
H_{\mu\nu}^{(i)} &=& \bar{u}(p_2) \gamma_{\mu} \gamma_5 u(-p_1)A_{2\gamma} \left[ \bar{u}(p_2) ( \tilde G_M \gamma_{\nu} - G_2 P_{\nu} ) u(-p_1) \right]^*\nn \\
&=& Tr \left[ (\hat p_2 +m ) \gamma_{\mu} \gamma_5 A_{2\gamma} (\hat p_1 -m) \frac{1}{2} (1-\gamma_5 \hat s_1) ( \tilde G_M^* \gamma_{\nu} -G_2^* P_{\nu} ) \right], \label{Himinicele}
\ea
which gives for the unpolarized part
\ba
H_{\mu\nu}^{(i)}(0) &=& \frac{1}{2}Tr \left[ (\hat p_2 +m ) \gamma_{\mu} \gamma_5 A_{2\gamma} (\hat p_1 -m) ( \tilde G_M^* \gamma_{\nu} -G_2^* P_{\nu} ) \right] \nn \\
&=& \frac{1}{2} A_{2\gamma} \tilde G_M^* Tr [ \gamma_5 \hat p_2 \gamma_{\mu} \hat p_1 \gamma_{\nu} ] = 2i A_{2\gamma} \tilde G_M^* \left< \mu \nu p_2 p_1 \right>. \label{Himini0}
\ea

\subsubsection{The polarized hadron tensor : $H_{\mu\nu}^{(i)}(s_1)$}

The polarized part of $H_{\mu\nu}^{(i)}$ follows from Eq. (\ref{Himinicele})
\ba
H_{\mu\nu}^{(i)}(s_1) &=& - \frac{1}{2}Tr \left[ (\hat p_2 +m ) \gamma_{\mu} \gamma_5 A_{2\gamma} (\hat p_1 -m) \gamma_5 \hat s_1 ( \tilde G_M^* \gamma_{\nu} -G_2^* P_{\nu} ) \right] \nn \\
&=& \frac{1}{2} A_{2\gamma} \Big[ Tr [ (\hat p_2 +m ) \gamma_{\mu} m \hat s_1 ( \tilde G_M^* \gamma_{\nu} -G_2^* P_{\nu})] \nn \\
&&+ Tr [ (\hat p_2 +m ) \gamma_{\mu} \hat p_1 \hat s_1 ( \tilde G_M^* \gamma_{\nu} -G_2^* P_{\nu})] \Big] \nn \\
&=& \frac{1}{2} A_{2\gamma} \Big[ - m^2 G_2^* P_{\nu} Tr [ \gamma_{\mu} \hat s_1 ] +m \tilde G_M^* Tr [ \hat p_2 \gamma_{\mu} \hat s_1 \gamma_{\nu}] \nn \\
&&- G_2^* P_{\nu} Tr [ \hat p_2 \gamma_{\mu} \hat p_1 \hat s_1 ] + m \tilde G_M^* Tr[ \gamma_{\mu} \hat p_1 \hat s_1 \gamma_{\nu} ] \Big]
\ea
and by applying the rules for calculating the traces we get
\ba
H_{\mu\nu}^{(i)}(s_1) &=& 2 A_{2\gamma} \Big[ - m^2 G_2^* P_{\nu} s_{1\mu} - G_2^* P_{\nu} (p_{2\mu} p_1 \cdot s_1 +p_{1\mu} p_2 \cdot s_1 - p_1 \cdot p_2 s_{1\mu})  \label{Himinipol}\\
&&+ m \tilde G_M^* (p_{2\mu} s_{1\nu} +p_{2\nu} s_{1\mu} - p_2 \cdot s_1 g_{\mu\nu} + p_{1\mu} s_{1\nu} - p_{1\nu} s_{1\mu} + p_1 \cdot s_1 g_{\mu\nu} ) \Big], \nn
\ea
where 
$$
s_1 \cdot p_1=0 \: ; \: s_1 \cdot p_2 = s_1 \cdot q,
$$
while $s_{1\mu}$ is polarization four-vector of the antinucleon. Using  Eq.  (\ref{idenhad}), expression (\ref{Himinipol}) can be simplified to
\ba
H_{\mu\nu}^{(i)}(s_1) &=& 2 A_{2\gamma} \Big[ 2m^2 (\tau-1) G_2^* P_{\nu} s_{1\mu} - G_2^* P_{\nu} p_{1\mu} s_1 \cdot q  - m \tilde G_M^* q \cdot s_1 g_{\mu\nu} \nn \\
&&+  m \tilde G_M^* (p_{2\mu} s_{1\nu} +p_{2\nu} s_{1\mu} + p_{1\mu} s_{1\nu} - p_{1\nu} s_{1\mu} ) \Big],
\ea
which can be rewritten in terms of the generalized Sachs FFs as
\ba 
H_{\mu\nu}^{(i)}(s_1) &=& m A_{2\gamma} \Big[ - 2q \cdot s_1 \tilde G_M^* g_{\mu\nu} - \frac{2q \cdot s_1}{m^2 (1-\tau)}(\tilde G_M-\tilde G_E)^*p_{1\mu} P_{\nu}  \label{Himini1} \\
&+& (\tilde G_M+\tilde G_E)^* (s_{1\mu} p_{2\nu} + s_{1\nu} p_{2\mu}) + (\tilde G_M - \tilde G_E)^* (s_{1\mu} p_{1\nu} + s_{1\nu} p_{1\mu}) \nn \\
&-& (\tilde G_M +\tilde  G_E)^* (s_{1\mu} p_{1\nu} - s_{1\nu} p_{1\mu}) - (\tilde G_M - \tilde G_E)^* (s_{1\mu} p_{2\nu} - s_{1\nu} p_{2\mu}) \Big], \nn
\ea
where we can distinguish two antisymmetric terms, three symmetric terms and the term proportional to $p_{1\mu} P_{\nu}$ ($2 p_{1\mu} P_{\nu}= p_{1\mu} P_{\nu}+p_{1\nu} P_{\mu}+p_{1\mu} P_{\nu}-p_{1\nu} P_{\mu}$). 
\section{Differential cross section }

The differential cross section can be written as the sum of unpolarized and polarized terms, corresponding to the different polarization states and polarization direction of the incident and scattered particles. In our case we consider just polarization of the outgoing antinucleon and longitudinal polarization of the incoming electron (with the degree of polarization $\lambda_e$).
\ba
\frac{d\sigma}{d\Omega}= \frac{d\sigma_{un}}{d\Omega}\left[ 1+ P_y\xi_y+\lambda_e P_x\xi_x + \lambda_e P_z\xi_z\right]. \label{difcrosec}
\ea
\subsection{Unpolarized differential cross section}
The unpolarized differential cross section can be written as
\begin{eqnarray*}
\frac{d\sigma_{un}}{d\Omega}=\frac{\alpha^2\beta}{4q^6} \left[ L_{\mu\nu}^{(v)}(0) H_{\mu\nu}^{(v)}(0) +2 Re (L_{\mu\nu}^{(i)}(0) H_{\mu\nu}^{(i)}(0))  \right]=\frac{\alpha^2\beta}{4q^2}D,
\end{eqnarray*}
where  $\beta=\sqrt{1-4m^2/q^2}$ is nucleon velocity in CMS and
\ba
D=\frac{1}{q^4}\left[ L_{\mu\nu}^{(v)}(0) H_{\mu\nu}^{(v)}(0) +2 Re (L_{\mu\nu}^{(i)}(0) H_{\mu\nu}^{(i)}(0) ) \right]. 
\ea

Let us calculate the first term of $D$. According to Eqs. (\ref{Lvmunu0}) and (\ref{Hvmunu0})
\ba
L_{\mu\nu}^{(v)}(0) H_{\mu\nu}^{(v)}(0) &=& \left[-q^2g_{\mu\nu} + 2\left(k_{1\nu}k_{2\mu}+k_{1\mu}k_{2\nu}\right)\right] \times \left[H_1 \tilde{g}_{\mu\nu} + H_2 P_{\mu} P_{\nu}\right] \nn \\
&=& -H_1 q^2 (4-\frac{q^2}{q^2}) - H_2 q^2 P^2 +4H_1 
\left(k_1\cdot k_2-\frac{(k_1\cdot q)(k_2\cdot q)}{q^2}\right) \nn \\
&&+ 4 H_2 (k_1\cdot P) (k_2\cdot P) ,
\ea
where $k_1\cdot k_2=k_1\cdot q=k_2\cdot q=q^2/2$ and 
\ba
P^2=\frac{(m^2-p_1\cdot p_2)}{2}=\frac{m^2-(E^2-\vec{p}_1 \cdot\vec{p}_2)}{2}=m^2-E^2=m^2(1-\tau),
\ea
where $E^2=q^2/4$ and $m^2=E^2-\vec{p}_1^{\:2}$.

Let us define a coordinate frame in CMS of the reaction
$e^+ + e^- \rightarrow N + \bar N$ in such a way that the $z$ axis is directed along the three-momentum of the antinucleon ($\vec{p}_1$). Therefore, the components of four-momenta can be written as
\ba
p_1=(E,0,0,|\vec{p}_1|) \: &;& \: k_1=(E,-|\vec{k}_1|\sin\theta,0,|\vec{k}_1|\cos\theta) \nn \\
p_2=(E,0,0,-|\vec{p}_1|)  \: &;& \: k_2=(E,|\vec{k}_1|\sin\theta,0,-|\vec{k}_1|\cos\theta) \nn \\
q=(2E,0,0,0) \: &;& \: P=(0,0,0,-|\vec{p}_1|), \label{CMSvectors}
\ea 
where $|\vec{k}_1|=E=m\sqrt{\tau}$, $|\vec{p}_1|=\sqrt{E^2-m^2}=m\sqrt{\tau -1}$ and $\theta$ is the angle between electron and detected antinucleon momenta.
These identities and definitions lead to
\ba
L_{\mu\nu}^{(v)}(0) H_{\mu\nu}^{(v)}(0) &=& 2 q^4 |\tilde G_M|^2 + 4 m^2 q^2 (|\tilde G_E|^2 - \tau |\tilde G_M|^2) \nn \\
&&- \frac{16}{\tau-1} (|\tilde G_E|^2 - \tau |\tilde G_M|^2) |\vec{k}_1|^2 |\vec{p}_1|^2 \cos^2 \theta \nn \\
&=& q^4 \left[ |\tilde G_M|^2 +\frac{1}{\tau} |\tilde G_E|^2 -\frac{1}{\tau} (|\tilde G_E|^2 - \tau |\tilde G_M|^2) \cos^2 \theta \right] \nn \\
&=& q^4 \left[ |\tilde G_M|^2 (1+\cos^2 \theta) +\frac{1}{\tau} |\tilde G_E|^2 \sin^2 \theta \right].
\ea
The second term of $D$ can be written according to Eqs. (\ref{Limini0}) and (\ref{Himini0}) as
\ba
L_{\mu\nu}^{(i)}(0) H_{\mu\nu}^{(i)}(0) &=& 2i \left< \mu \nu k_2 k_1 \right> \times 2i A_{2\gamma} \tilde G_M^* \left<\mu \nu p_2 p_1 \right>, \nn
\ea
which can be written as (see appendix)
\ba
L_{\mu\nu}^{(i)}(0) H_{\mu\nu}^{(i)}(0) &=& -4 A_{2\gamma} \tilde G_M^* 2 \left((k_2\cdot p_1)  (k_1\cdot p_2)-(k_2\cdot p_2)(k_1\cdot p_1)\right) \nn \\
&=& q^4 \left[-2 A_{2\gamma} \tilde G_M^*  \frac{1}{\tau} \sqrt{\tau(\tau-1)} \cos \theta \right].
\ea

Finally we get the following expression for $D$
\ba
D=|\tilde G_M|^2 (1+\cos^2 \theta) + \frac{1}{\tau} |\tilde G_E|^2 \sin^2 \theta - \frac{4}{\tau} \sqrt{\tau(\tau-1)} \cos \theta Re \tilde G_M A_{2\gamma}^*. 
\ea

\subsubsection{$2\gamma$ mechanism and the unpolarized cross section}

To separate the effects due to the Born and TPE contributions, let us single 
out the dominant contribution and define the following decompositions of the 
amplitudes 
\begin{eqnarray}
\tilde G_{M}(q^2, t)&=&G_{M}(q^2)+\Delta G_{M}(q^2, t), \nonumber\\
\tilde G_{E}(q^2, t)&=&G_{E}(q^2)+\Delta G_{E}(q^2, t). 
\label{eq:eq6}
\end{eqnarray}
$\Delta G_{M}(q^2, t),$ $\Delta G_{E}(q^2, t),$ and $A_{2\gamma}(q^2, t)$ are of the 
order of $\sim \alpha $, while $G_{M}(q^2)$ and $G_{E}(q^2)$ are of the order of $\sim \alpha^0.$ 

Symmetry properties of the amplitudes
with respect to the $\cos\theta \to -\cos\theta $ transformation can be derived in model independent way, from the $ C $
invariance of the $1\gamma \otimes 2\gamma $ mechanism.

To prove this, let us consider, in addition to C-invariance, crossing symmetry, which allows to connect the matrix elements for the cross-channels: 
$e^-(k_1)+N(p_1)\to e^-(k_2)+N(p_2)$, in $s$--channel, and $e^++e^-\to N+\overline{N}$, in $t$--channel. The transformation from $s$- to $t$-channel can be realized by the following substitution:
$$k_2\to -k_2,~p_1\to -p_1.$$
and for the invariant variables:
$$s=(k_1+p_1)^2\to (k_1-p_1)^2,~Q^2=-(k_1-k_2)^2\to -(k_1+k_2)^2=-t.$$

Crossing symmetry states that the same three amplitudes $\tilde G_E(s,Q^2)$, $\tilde G_M(s,Q^2)$  and $A_{2\gamma}(s,Q^2)$ describe the two channels, when the variables $s$ and $Q^2$ scan the physical region of the corresponding channels. So, if $t\ge 4m^2$ and $-1\le\cos\theta\le 1$ ($\theta$ is the angle of the proton production with respect to the electron three-momentum, in the center of mass (CMS) for $e^++e^-\to N+\overline{N}$), the amplitudes $\tilde G_E(t,\cos\theta)$, $\tilde G_M(t,\cos\theta)$, and $A_{2\gamma}(t,\cos\theta)$, describe the process $e^++e^-\to p+\overline{p}$.

The C-invariance of the electromagnetic hadron interaction and the corresponding selection rules can be  applied to the annihilation channel and this allows to find specific properties for one and two photon exchanges. Moreover, on the basis of the crossing symmetry, it is possible to transform in a transparent way these properties for the different observables in $eN$-elastic scattering.

To illustrate this, let us consider firstly the one-photon mechanism for $e^++e^-\to p+\overline{p}$. The conservation of the total angular momentum ${\cal J}$  allows one value, ${\cal J}=1$ , and the quantum numbers of the photon: ${\cal J}^P=1^-$, $C=-1$. The selection rules with respect to the C and P-invariance allow two states for 
$e^+e^-$ (and $p\overline{p}$):
\begin{equation}
S=1,~\ell=0 \mbox{~and~} S=1,~\ell=2\mbox{~with~} {\cal J}^P=1^-,
\label{eq:tran}
\end{equation}
where $S$ is the total spin and $\ell$ is the orbital angular momentum. As a result the $\theta$-dependence of the cross section for $e^++e^-\to p+\overline{p}$, in the one-photon exchange mechanism is:
\begin{equation}
\displaystyle\frac{d\sigma}{d \Omega}(e^++e^-\to p+\overline{p})\simeq a(t)+b(t)\cos^2\theta, 
\label{eq:sig}
\end{equation}
where $a(t)$ and $b(t)$ are definite quadratic contributions of $G_{Ep}(t)$ and 
$G_{Mp}(t)$, $a(t),~b(t)\ge 0$ at $t\ge 4m^2$.

Using the kinematical relations:
\begin{equation}
\cos^2\theta=\displaystyle\frac{1+\epsilon }{1-\epsilon}=
\displaystyle\frac{\cot^2{\theta_e/2}}{1+\tau}+1
\label{eq:cot}
\end{equation}
between the variables in the CMS of $e^++e^-\to p+\overline{p}$ and in the LAB system for $e^-+p\to e^-+p$, it appears clearly that the one-photon mechanism generates a linear $\epsilon$-dependence (or  $\cot^2{\theta_e/2}$) of the Rosenbluth differential cross section for elastic $eN$-scattering in Lab system.

Let us consider now the $\cos\theta$-dependence of the $1\gamma\bigotimes 2\gamma$-interference contribution to the differential cross section of  $e^++e^-\to p+\overline{p}$. The spin and parity of the $2\gamma$-states 
is not fixed, in general, but only a positive value of C-parity, $C(2\gamma)=+1$, is allowed.
An infinite number of  states with different quantum numbers can contribute, and their relative role is determined by the dynamics of the process $\gamma^*+\gamma^*\to  p+\overline{p}$, with both virtual photons.

But the $\cos\theta$-dependence of the contribution to the differential cross section for the $1\gamma\bigotimes 2\gamma$-interference can be predicted on the basis of its C-odd nature:
\begin{equation}
\displaystyle\frac{d\sigma^{(int)}}{d \Omega}(e^++e^-\to p+\overline{p})=\cos\theta[c_0(t)+c_1(t)\cos^2\theta+c_2(t)\cos^4\theta+...],
\label{eq:sig3}
\end{equation}
where $c_i(t)$, $i=0,1..$ are real coefficients, which are functions of $t$,  only. This odd $\cos\theta$-dependence is essentially different from the even $\cos\theta$-dependence of the cross section for the one-photon approximation.
It is therefore incorrect to approximate the interference contribution to the differential cross section
(\ref{eq:sig3}) by a linear function in $\cos^2\theta$, because it is in contradiction with the C-invariance of hadronic electromagnetic interaction.
Such approximation can be done only when all coefficients $c_i(t)$ vanish, i.e. in absence of $1\gamma\bigotimes 2\gamma$-interference!

Therefore $C$-invariance predict the following symmetry of the two-photon contribution to the amplitudes in $e^+e^-$ annihilation:
\begin{equation}
\Delta G_{M,E}(\cos\theta )=-\Delta G_{M,E}(-\cos\theta ),
\ A_{2\gamma}(\cos\theta )=A_{2\gamma}(-\cos\theta ).
\label{eq:eq14}
\end{equation}
Let us consider the situation when the experimental apparatus does not distinguish the nucleon
from the antinucleon. Then we measure the following sum of the differential cross sections
$$\frac{d\sigma_+}{d\Omega}=\frac{d\sigma}{d\Omega}(\cos\theta )+\frac{d\sigma}{d\Omega}(-\cos\theta ). $$
We can stress, using the properties (\ref{eq:eq14}), that this quantity does not depend on the TPE
terms. 

Note also that the TPE terms do not contribute to the total cross section of the reaction
$ e^++e^-\rightarrow N+\bar N$,  which can be written as
\begin{equation}\label{eq:eq15}
\sigma _t(q^2)=\frac {4\pi }{3}\frac {\alpha ^2\beta }{q^2}\left [|G_M(q^2)|^2+\frac{1}{2\tau }|G_E(q^2)|^2
\right ].
\end{equation}
On the other hand, the relative contribution of TPE mechanism is enhanced in the following
angular asymmetry 
\begin{equation}\label{eq:eq16}
A_{2\gamma}(q^2,\theta_0)=\frac {\sigma (q^2,\theta_0)-\sigma (q^2,\pi -\theta_0)}
{\sigma (q^2,\theta_0)+\sigma (q^2,\pi -\theta_0)},
\end{equation}  
where the quantities $\sigma (q^2,\theta_0)$ and $\sigma (q^2,\pi -\theta_0)$ are defined as follows
$$\sigma (q^2,\theta_0)=\int_0^{\theta_0}\frac{d\sigma }{d\Omega}(q^2,\theta )d\Omega , \ \
\sigma (q^2,\pi -\theta_0)=\int_{\pi -\theta_0}^{\pi }\frac{d\sigma }{d\Omega}(q^2,\theta )d\Omega . $$
Using the symmetry relations (\ref{eq:eq14}) one  can obtain for the asymmetry $A_{2\gamma}(q^2, \theta_0)$ the  following expression
\begin{eqnarray}
A_{2\gamma}(q^2, \theta_0)&=&\frac{2}{d}\int_0^{\theta_0}d\cos\theta \Bigg [(1+\cos^2\theta )ReG_M(q^2)\Delta G_M^*(q^2,\cos\theta )
\nonumber \\
&&+\frac{\sin^2\theta }{\tau }ReG_E(q^2)\Delta G_E^*(q^2,\cos\theta ) \nonumber \\
&&-\frac{2}{\tau }\sqrt{\tau (\tau -1)}\cos\theta ReG_M(q^2)A^*(q^2,\cos\theta )\Bigg ],
\label{eq:eq17}
\end{eqnarray}
where the quantity $d$ is
$$d=\frac{1-x_0}{3}\Bigg [(4+x_0+x_0^2)|G_M|^2+\frac{1}{\tau }(2-x_0-x_0^2)|G_E|^2 \Bigg ], \ \ x_0=\cos\theta_0 . $$

The TPE contributions can be removed considering the sum of the quantities $\sigma (q^2,\theta_0)$
and $\sigma (q^2,\pi -\theta_0)$. As a result we have
\begin{equation}
\Sigma (q^2, \theta_0)=\sigma (q^2,\theta_0)+\sigma (q^2,\pi -\theta_0)=\frac{\pi \alpha^2}{q^2}\beta d 
\label{eq:eq18}
\end{equation}
always neglecting the terms of the order of $\alpha ^2$ with respect to the leading ones.

\subsection{Single spin polarization observables, antiproton polarization $P_y$}

$P_y$ is a single-spin polarization observable, which appears in the Born approximation in the $e^- + e^+ \rightarrow N + \bar{N}$ process with one polarized particle - the antinucleon ($\bar N$), which is polarized along the $y$-axis. It is shown below, that this observable doesn't depend on polarization of electron. Polarization of antinucleon along $y$-axis means, that its polarization unit vector $\vec{\xi}$ has only $y$-component ($\vec{\xi}=(0,1,0)$). This leads to following properties of antinucleon $s_{1y}$ (\ref{4pol}), (\ref{CMSvectors})
\ba
\vec p_1 \cdot  \vec{\xi} = 0 \Rightarrow s_{10}=0 \: ; \: \vec s_{1y} = \vec{\xi} = (0,1,0).  \label{sy}
\ea

The general expression for $P_y$ is
\ba
P_y &=& \frac{\alpha^2\beta}{4q^6} \left[ L_{\mu\nu}^{(v)} H_{\mu\nu}^{(v)}(s_{1y}) +2 Re (L_{\mu\nu}^{(i)} H_{\mu\nu}^{(i)}(s_{1y}) ) \right] \Big/ \frac{d\sigma_{un}}{d\Omega} \nn \\
&=& \frac{1}{D q^4} \left[ L_{\mu\nu}^{(v)} H_{\mu\nu}^{(v)}(s_{1y}) +2 Re (L_{\mu\nu}^{(i)} H_{\mu\nu}^{(i)}(s_{1y}) ) \right], \label{Pydef}
\ea
which can be divided into two parts - with unpolarized electron and with polarized electron
\ba
P_y &=& \frac{1}{D q^4} \left[ L_{\mu\nu}^{(v)}(0) H_{\mu\nu}^{(v)}(s_{1y}) +2 Re (L_{\mu\nu}^{(i)}(0) H_{\mu\nu}^{(i)}(s_{1y}) ) \right] \nn \\
&+& \frac{1}{D q^4} \left[ L_{\mu\nu}^{(v)}(S) H_{\mu\nu}^{(v)}(s_{1y}) +2 Re (L_{\mu\nu}^{(i)}(S) H_{\mu\nu}^{(i)}(s_{1y}) ) \right]. \nn
\ea

Firstly we will prove that longitudinally polarized electron doesn't contribute to the polarization observable $P_y$. The first term of the polarized electron part equals (\ref{Lvmini1}), (\ref{Hvmini1}) for $\lambda_e=1$:
\ba
L_{\mu\nu}^{(v)}(S) H_{\mu\nu}^{(v)}(s_{1y}) &=&  2i \left\langle \mu \nu k_1 q \right\rangle \times \frac{2}{m(\tau-1)} \Big[ i m^2 (\tau - 1) |\tilde G_M|^2 \left< \mu \nu q s_{1y} \right> \nn \\
&&+ i Re(\tilde G_M (\tilde G_E-\tilde G_M)^*) ( P_{\mu} \left< \nu p_2 p_1 s_{1y} \right> - P_{\nu} \left< \mu p_2 p_1 s_{1y} \right> ) \nn \\
&&+ Im (\tilde G_M \tilde G_E^*) (P_{\mu} \left< \nu p_2 p_1 s_{1y} \right> + P_{\nu} \left< \mu p_2 p_1 s_{1y} \right> ) \Big] . \nn
\ea
The lepton tensor is antisymmetric, therefore its product with the third (symmetric) part of the hadron tensor vanishes. The first product is proportional to
\ba
\left\langle \mu \nu k_1 q \right\rangle \times \left< \mu \nu q s_{1y} \right> = 2( k_1\cdot s_{1y} ~ q^2 - k_1\cdot q ~ s_{1y}\cdot q )= 0, \label{step1}
\ea
where we used (\ref{CMSvectors}, \ref{sy}) $k_1.s_{1y}=s_{1y}.q=0$.  The second product is proportional to
\ba
\left< \mu \nu k_1 q \right> &\times& ( P_{\mu} \left< \nu p_2 p_1 s_{1y} \right> - P_{\nu} \left< \mu p_2 p_1 s_{1y} \right> ) = - 2 \left\langle \nu P k_1 q \right\rangle \left< \nu p_2 p_1 s_{1y} \right>  \nn \\
&=& -2\big[ P\cdot p_2 (k_1\cdot  s_{1y} ~ q\cdot  p_1 - k_1\cdot  p_1 ~ q\cdot s_{1y}) +P\cdot p_1 ( k_1\cdot p_2 ~ q\cdot  s_{1y} - k_1\cdot s_{1y} ~ q\cdot p_2) \nn \\
&&+ P\cdot s_{1y} (k_1 \cdot  p_1 ~ q\cdot p_2 - k_1\cdot p_2 ~ q\cdot p_1)\big]=0, \label{step2}
\ea 
where again Eqs. (\ref{CMSvectors}, \ref{sy}) was used ($P\cdot s_{1y}=q\cdot s_{1y}=k_1\cdot s_{1y}=0$).

The second term of the polarized electron part of $P_y$ (\ref{Limini1}), (\ref{Himini1})
\ba
L_{\mu\nu}^{(i)}(S) H_{\mu\nu}^{(i)}(s_{1y}) &=& \left[ q^2 g_{\mu\nu} - 2 (k_{2\mu} k_{1\nu} + k_{2\nu} k_{1\mu}) \right] \times 
\nn \\
&&\times m A_{2\gamma} \Big[ - 2q\cdot s_{1y}\tilde G_M^* g_{\mu\nu} - \frac{2q\cdot s_{1y}}{m^2 (1-\tau)}(\tilde G_M-\tilde G_E)^*p_{1\mu} P_{\nu}  \nn \\
&&+ (\tilde G_M+\tilde G_E)^* (s_{1y\mu} p_{2\nu} + s_{1y\nu} p_{2\mu}) + (\tilde G_M - \tilde G_E)^* (s_{1y\mu} p_{1\nu} + s_{1y\nu} p_{1\mu}) \nn \\
&&- (\tilde G_M + \tilde G_E)^* (s_{1y\mu} p_{1\nu} - s_{1y\nu} p_{1\mu}) - (\tilde G_M - \tilde G_E)^* (s_{1y\mu} p_{2\nu} - s_{1y\nu} p_{2\mu}) \Big], \nn
\ea	
where taking into account that $q\cdot s_{1y}=0$ and that the product of a symmetric tensor ($L_{\mu\nu}^{(i)}$) and an antisymmetric tensor is zero leads to
\ba
L_{\mu\nu}^{(i)}(S) H_{\mu\nu}^{(i)}(s_{1y})&=&2 q^2 m A_{2\gamma} \left[ (\tilde G_M + \tilde G_E )^*  s_{1y}\cdot p_2 + (\tilde G_M - \tilde G_E )^* s_{1y}\cdot p_1 \right ] \nn \\
&&- 4 m A_{2\gamma} \big( (\tilde G_M +\tilde  G_E)^*  (k_2\cdot s_{1y} ~ k_1\cdot p_2 + k_1\cdot s_{1y} ~ k_2\cdot p_2) \nn \\
&&+ (\tilde G_M - \tilde G_E)^* (k_2.s_{1y} ~ k_1\cdot p_1 + k_1\cdot s_{1y} ~ k_2\cdot p_1) \big)=0, \label{step3}
\ea
while $k_2\cdot s_{1y}=k_1\cdot s_{1y}=p_1\cdot s_{1y}=p_2\cdot s_{1y}=0$.

Therefore, the polarization observable $P_y$ depends only on the unpolarized part
\ba
P_y &=& \frac{1}{D q^4} \left[ L_{\mu\nu}^{(v)}(0) H_{\mu\nu}^{(v)}(s_{1y}) +2 Re (L_{\mu\nu}^{(i)}(0) H_{\mu\nu}^{(i)}(s_{1y}) ) \right].   \label{Pypred}
\ea
With the help of Eqs. (\ref{Lvmunu0}, \ref{Hvmini1}), the first term is equal to 
\ba
L_{\mu\nu}^{(v)}(0) H_{\mu\nu}^{(v)}(s_{1y})  &=& \big[ -q^2g_{\mu\nu} + 2\left(k_{1\nu}k_{2\mu}+k_{1\mu}k_{2\nu}\right)\big] \nn \\
&&\times \frac{2}{m(\tau-1)} \Big[ i m^2 (\tau - 1) |\tilde G_M|^2 \left< \mu \nu q s_{1y} \right> \nn \\
&&+ i Re(\tilde G_M (\tilde G_E-\tilde G_M)^*) ( P_{\mu} \left< \nu p_2 p_1 s_{1y} \right> - P_{\nu} \left< \mu p_2 p_1 s_{1y} \right> ) \nn \\
&&+ Im (\tilde G_M \tilde G_E^*) (P_{\mu} \left< \nu p_2 p_1 s_{1y} \right> + P_{\nu} \left< \mu p_2 p_1 s_{1y} \right> ) \Big]. \nn
\ea
As $L_{\mu\nu}^{(v)}(0)$ is a symmetric tensor it gives non-zero product only with the last (symmetric) part of $H_{\mu\nu}^{(v)}(s_{1y})$
\ba
L_{\mu\nu}^{(v)}(0) H_{\mu\nu}^{(v)}(s_{1y})  &=& \big[ -q^2g_{\mu\nu} + 2\left(k_{1\nu}k_{2\mu}+k_{1\mu}k_{2\nu}\right)\big] \nn \\
&&\times \frac{2}{m(\tau-1)} \left [ Im ( \tilde G_M  \tilde G_E^*) (P_{\mu} \left< \nu p_2 p_1 s_{1y} \right> + P_{\nu} \left< \mu p_2 p_1 s_{1y} \right> ) \right ] \nn \\
&=& \frac{2 Im (\tilde G_M \tilde G_E^*)}{m(\tau-1)} \left [ - q^2 \left< p_2 p_2 p_1 s_{1y} \right> + q^2 \left< p_1 p_2 p_1 s_{1y} \right> \right . \nn \\
&&+ \left . 4 k_1\cdot P \left< k_2 p_2 p_1 s_{1y} \right> + 4 k_2\cdot P \left< k_1 p_2 p_1 s_{1y} \right>\nn \right ],
\ea
where $\left< p_2 p_2 p_1 s_{1y} \right>=\left< p_1 p_2 p_1 s_{1y} \right>=0$, because they are antisymmetric with two equal components. 

The computation of $\left< k_2 p_2 p_1 s_{1y} \right>$ and $\left< k_1 p_2 p_1 s_{1y} \right>$ is more complicated and for the first time we will make it in detail. Let us recall the definition of 
\ba
\left< k_2 p_2 p_1 s_1 \right>=\varepsilon_{\mu\nu\rho\sigma} k_{2\mu} p_{2\nu} p_{1\rho} s_{1y\sigma},~ \mu,\nu,\rho,\sigma=0...3 . \label{fancybrackets}
\ea
We get non-zero result only if indices $\mu,\nu,\rho,\sigma$ are different from each other (due to antisymmetric $\varepsilon$-tensor property) and components $k_{2\mu}, p_{2\nu}, p_{1\rho}, s_{1y\sigma}$ are non-zero for the given index.  Notice, that in case of $P_y$ polarization four-vector $s_{1y}$ has only one ($y$) non-zero component (\ref{sy}), so in the equation (\ref{fancybrackets}) $\sigma=2$. On the other hand four-momentum $k_2$ is the only one with non-zero $x$-component and therefore $\mu=1$. At last $p_1$ and $p_2$ have two nonzero components, what leads into (with $\varepsilon_{1230}=1$) 
\ba
\varepsilon_{1\nu\rho2} k_{2x} p_{2\nu} p_{1\rho} s_{1yy}&=&-\varepsilon_{1032} k_{2x} p_{20} p_{1z} -\varepsilon_{1302} k_{2x} p_{2z} p_{10}  \nn \\
&=&k_{2x} p_{20} p_{1z} -k_{2x} p_{2z} p_{10} = 2 E |\vec k_1| |\vec p_1| \sin \theta \nn \\
&=& \frac{q^2}{2}m\sqrt{\tau-1} \sin\theta.
\ea
For the $\left< k_1 p_2 p_1 s_{1y} \right>$ we obtain a similar result
$$
\left< k_1 p_2 p_1 s_{1y} \right>=-\frac{q^2}{2}m\sqrt{\tau-1} \sin\theta
$$
and according to Eq. (\ref{CMSvectors}) $k_1\cdot P=-k_2\cdot P=m^2 \sqrt{\tau(\tau-1)} \cos \theta$, which all together give a result
\ba
L_{\mu\nu}^{(v)}(0) H_{\mu\nu}^{(v)}(s_{1y})  &=& \frac{8}{m(\tau-1)} Im (\tilde G_M \tilde G_E^*) q^2 m^3 (\tau-1) \sqrt{\tau} \sin \theta \cos \theta \nn \\
&=& 8 m^2 q^2 Im (\tilde G_M \tilde G_E^*) \sqrt{\tau} \sin \theta \cos \theta. \label{Pyprvy}
\ea
With the help of Eqs. (\ref{Limini0}, \ref{Himini1}), the second term of Eq.  (\ref{Pypred}) is equal to 
\ba
L_{\mu\nu}^{(i)}(0) H_{\mu\nu}^{(i)}(s_{1y}) &=& 2i\left\langle \mu \nu k_2 k_1 \right\rangle  m A_{2\gamma} 
\left [ - 2q\cdot s_{1y} \tilde G_M^* g_{\mu\nu} - \frac{2q\cdot s_{1y}}{m^2 (1-\tau)}(\tilde G_M-\tilde G_E)^*p_{1\mu} P_{\nu}\right .  \nn \\
&&+ (\tilde G_M+\tilde G_E)^* (s_{1y\mu} p_{2\nu} + s_{1y\nu} p_{2\mu}) + (\tilde G_M - \tilde G_E)^* (s_{1y\mu} p_{1\nu} + s_{1y\nu} p_{1\mu}) \nn \\
&&- \left .(\tilde G_M + \tilde G_E)^* (s_{1y\mu} p_{1\nu} - s_{1y\nu} p_{1\mu}) - (\tilde G_M - \tilde G_E)^* (s_{1y\mu} p_{2\nu} - s_{1y\nu} p_{2\mu}) \right ], \nn
\ea 
where $q\cdot s_{1y}=0$ and the product of the antisymmetric $L_{\mu\nu}^{(i)}(0)$  with the symmetric parts of $H_{\mu\nu}^{(i)}(s_{1y})$
vanishes:
\ba
&&L_{\mu\nu}^{(i)}(0) H_{\mu\nu}^{(i)}(s_{1y}) \nn \\
&&= 
- 2i m A_{2\gamma} \left\langle \mu \nu k_2 k_1 \right\rangle \Big[(\tilde G_M + \tilde G_E)^* (s_{1y\mu} p_{1\nu} - s_{1y\nu} p_{1\mu})
+ (\tilde G_M - \tilde G_E)^* (s_{1y\mu} p_{2\nu} - s_{1y\nu} p_{2\mu}) \Big] \nn \\
&&= - 4i m A_{2\gamma} \Big[ \tilde G_M^* ( \left\langle s_{1y} p_1 k_2 k_1 \right\rangle + \left\langle s_{1y} p_2 k_2 k_1 \right\rangle)+ \tilde G_E^* ( \left\langle s_{1y} p_1 k_2 k_1 \right\rangle - \left\langle s_{1y} p_2 k_2 k_1 \right\rangle) \Big] \nn
\ea
and
\ba
\left\langle s_{1y} p_1 k_2 k_1 \right\rangle+ \left\langle s_{1y} p_2 k_2 k_1 \right\rangle &=&\left\langle s_{1y} q k_2 k_1 \right\rangle = 0 \nn \\
\left\langle s_{1y} p_1 k_2 k_1 \right\rangle - \left\langle s_{1y} p_2 k_2 k_1 \right\rangle &=&-2\left\langle s_{1y} P k_2 k_1 \right\rangle.  \nn
\ea
Therefore
\ba
L_{\mu\nu}^{(i)}(0) H_{\mu\nu}^{(i)}(s_{1y}) &=& 8 i m A_{2\gamma}  \tilde G_E^* \left\langle s_{1y} P k_2 k_1 \right\rangle, \nn
\ea
where (similar to previous derivation)
\ba
\varepsilon_{23\mu\nu}  s_{yy} P_z k_{2\mu} k_{1\nu} &=& -\varepsilon_{2301} (-|\vec p_1|) k_{20} k_{1x} - \varepsilon_{2310} (-|\vec p_1|) k_{2x} k_{10} \nn \\
&=& 2 m^3 \tau \sqrt{\tau-1} \sin \theta. \nn
\ea
So the second term of Eq. (\ref{Pypred}) is
\ba
L_{\mu\nu}^{(i)}(0) H_{\mu\nu}^{(i)}(s_{1y}) = 16 i m^4 A_{2\gamma}  \tilde G_E^* \tau \sqrt{\tau-1} \sin \theta \label{Pydruhy}
\ea
and finally for $P_y$ (\ref{Pyprvy}, \ref{Pydruhy}) we get
\ba
P_y &=& \frac{2\sin \theta }{D\sqrt{\tau}} \Big[ Im (\tilde G_M \tilde G_E^*)  \cos \theta +\sqrt{\frac{\tau-1}{\tau}} Re [ i  A_{2\gamma}  \tilde G_E^* ] \Big] \nn \\
&=&  \frac{2\sin \theta }{D\sqrt{\tau}} \Big[ Im (\tilde G_M \tilde G_E^*)  \cos \theta + \sqrt{\frac{\tau-1}{\tau}} Im [ A_{2\gamma}^*  \tilde G_E ] \Big]. 
\ea
The polarization of the outgoing antinucleon in this case is determined
by the polarization component which is perpendicular to the reaction plane.
\subsubsection{$2\gamma$ mechanism and the single spin polarization $P_y$}

The polarization $P_y$, being T--odd quantity, does not vanish even in the
one--photon--exchange approximation due to the complexity of the nucleon FFs
in the TL region (to say more exactly, due to the non--zero difference
of the phases of these FFs). This is principal difference with the elastic
electron--nucleon scattering.

In the Born approximation this polarization becomes equal to zero at the
scattering angle $\theta = 90^0$ (as well at $\theta = 0^0$ and $180^0$).
The presence of the TPE contributions leads to a non--zero value of the
polarization at this angle and it is determined by a simple expression
$$P_y(90^0)=2\frac{\sqrt{\tau -1}}{\tau\bar D}ImG_EA_{2\gamma}^*,
\ \ \bar D=D(\theta =90^0). $$
Here the function $A_{2\gamma}$ is also calculated at the value $\theta =90^0$.
This quantity expected to be small due to the fact that it is determined by
the interference of the one--photon and two--photon exchange amplitudes and
should be of the order of $\alpha $. The measurement of this polarization at
$\theta = 90^0$ contains information about the TPE contribution and its
behavior as a function of $q^2.$

In the threshold region we can conclude that in the Born approximation this polarization vanishes, 
due to the relation $G_E=G_M$ which is valid at the threshold. The TPE contributions induces a non zero polarization, which is determined by a simple formula
$$P_y^{th}(\theta )=\frac{\sin 2\theta }{D^{th}}ImG_N(
\Delta G_E-\Delta G_M)^*. $$
Note that, at threshold, this polarization can still vanish if
$\Delta G_E=\Delta G_M$. In this case the differential cross section does not contain any explicit dependence on the angular variable $\theta $.
In the general case, the amplitudes $\Delta G_{E,M}$ depend on the $\theta $
variable. The effect of the TPE contributions for the polarization at an arbitrary scattering angle is expected to increase as $q^2$ increase, as the TPE amplitudes decrease more slowly with $q^2$ in comparison with the nucleon FFs.

Using the properties of the TPE amplitudes with respect to the 
$\cos\theta \to -\cos\theta $ transformation, one can remove the contributions of
the TPE effects by constructing the following quantities. Let us introduce the terms
$P_y(q^2,\theta_0)$ and $P_y(q^2,\pi -\theta_0)$, which are integrals of the 
polarization $P_y(q^2,\theta )$ over the angular regions connected by 
the above  mentioned transformation
$$P_y(q^2,\theta_0)=\int_0^{\theta_0}P_y(q^2,\theta )d\Omega , ~P_y(q^2,\pi -\theta_0)=\int_{\pi -\theta_0}^{\pi }P_y(q^2,\theta )d\Omega . $$ 
Let us calculate the sum and the difference of these two quantities. At the first order of the coupling constant $\alpha $, we obtain
\begin{eqnarray}
D^P(q^2,\theta_0)&=&P_y(q^2,\theta_0)-P_y(q^2,\pi -\theta_0)=\nonumber \\
&&=-\frac{8\pi R}{\sqrt{\tau }}(1-\frac{R^2}{\tau })^{-(3/2)}
\sin(\delta_{M}-\delta_{E})\Bigg
[\sqrt{z}+\frac{1}{\sqrt{2}}ln \left |\frac{\sqrt{z}-\sqrt{2}}{\sqrt{z}+\sqrt{2}}\right |\Bigg ], 
\label{eq:eq21}
\end{eqnarray}
where 
$$R=\frac{|G_E|}{|G_M|}, \ \ z=(1-x_0^2)(1-\frac{R^2}{\tau }), $$ 
and $\delta_M(\delta_E)$  is the phase of the complex FF $G_M(G_E)$. We can 
see that, in this approximation, the quantity $D^P$ does not depend on the TPE 
contribution. So, the phase difference of FFs can be correctly determined from this quantity, if the ratio $R$ is known. 

Let us consider the ratio of the function $\Sigma (q^2,\theta_0)$, Eq. (\ref{eq:eq18}), calculated at two values of $\theta_0$: 
$$\frac{\Sigma (q^2,\theta_1)}{\Sigma (q^2,\theta_2)}=\frac{1-x_1}{1-x_2}\cdot  
 \frac{4+x_1+x_1^2+\frac{1}{\tau }(2-x_1-x_1^2)R^2}{4+x_2+x_2^2+\frac{1}{\tau }(2-x_2-x_2^2)R^2}, 
 \ \  x_i=\cos\theta_i, \ i=1,2 $$
This ratio allows to determine $R$, minimizing systematic errors.

The magnitude of the TPE contribution to the polarization $P_y$, integrated over the considered angular region, can be obtained from the sum of the quantities
 introduced above
\begin{eqnarray}
\Sigma ^P(q^2,\theta_0)&=&P_y(q^2,\theta_0)+P_y(q^2,\pi -\theta_0) \nn \\
&=&\frac{8\pi }{\sqrt{\tau }}\int_0^{\theta_0}d\cos\theta \frac{sin\theta }{D_B}\Bigg 
\{\cos\theta Im(G_M\Delta G^*_E-G_E\Delta G^*_M) \nonumber \\
&&-2\frac{\cos\theta }{D_B}ImG_MG^*_E\Bigg [(1+\cos^2\theta )ReG_M\Delta G^*_M 
+\frac{\sin^2\theta }{\tau }ReG_E\Delta G^*_E\Bigg ] \nonumber \\
&&-\sqrt{\frac{\tau -1}{\tau }}\Bigg [ImG_EA_{2\gamma}^*+4\frac{\cos^2\theta }{D_B}
ImG_MG^*_EReG_MA^*_{N}\Bigg ]\Bigg \}, \label{eq:eq22}
\end{eqnarray} 
where  
$$D_B=(1+\cos^2\theta )|G_M|^2+\frac{\sin^2\theta }{\tau }|G_E|^2. $$
\section{Double spin polarization observables}

\subsection{The  component $P_x$}
$P_x$ is a double-spin polarization observable: the polarization of the incoming electron is necessary, in order to obtain a polarization of the outgoing antinucleon along the $x$-axis. The definition of the polarization observable $P_x$ is similar to $P_y$ (\ref{Pydef})
\ba
P_x = \frac{1}{D q^4} \left[ L_{\mu\nu}^{(v)} H_{\mu\nu}^{(v)}(s_{1x}) +2 Re (L_{\mu\nu}^{(i)} H_{\mu\nu}^{(i)}(s_{1x}) ) \right],
\ea
where, according to definition (\ref{4pol}), the four-vector $s_{1x}$ is 
\ba
\vec \xi = (1,0,0) \Rightarrow s_{x0}=0 \: ; \: \vec s_{1x} =(1,0,0).
\ea
For the derivation of $P_x$ we can use the same arguments as for $P_y$, with the following specificities:
\begin{itemize}
\item $k_1$, $k_2$ are not perpendicular to $s_{1x}$ and $k_1\cdot s_{1x}=-k_2\cdot s_{1x}=|\vec k_1| \sin \theta$. 
\item the fully contracted terms $\left<....\right>$, which contain only $s$, $k_1$, $k_2$, $p_1$, $p_2$, $q$, $P$ are vanishing, because these four-vectors have zero $y$-component.
\end{itemize}
The first property can be used in steps (\ref{step1}, \ref{step2}, \ref{step3}) and, as a consequence, the polarized electron part of $P_x$ is not vanishing. The second property can be used in the derivation of unpolarized electron part of $P_x$, where similarly to $P_y$, the only 'non-zero' terms are proportional to $\left<....\right>$ terms, which are zero for $P_x$. Therefore unpolarized electron process doesn't contribute to $P_x$.

Let us repeat steps (\ref{step1}, \ref{step2}) for $P_x$
\ba
\left\langle \mu \nu k_1 q \right\rangle \times \left< \mu \nu q s_{1x} \right> = 2(k_1\cdot s_{1x} ~ q^2 - k_1\cdot q ~ s_{1x} \cdot q) = 2 k_1\cdot s_{1x}  ~ q^2 = 2 m \sqrt{\tau} q^2 \sin \theta, \nn
\ea
where $s_{1x}\cdot q=0$. And
\ba
\left< \mu \nu k_1 q \right> &\times& ( P_{\mu} \left< \nu p_2 p_1 s_{1x} \right> - P_{\nu} \left< \mu p_2 p_1 s_{1x} \right> ) = -2\big[ P\cdot p_2 (k_1\cdot s_{1x} ~ q\cdot p_1 - k_1\cdot  p_1 ~ q\cdot s_{1x}) \nn \\
&&+ P\cdot p_1 ( k_1\cdot p_2 ~ q\cdot s_{1x} - k_1\cdot s_{1x} ~ q\cdot p_2) + P\cdot s_{1x} (k_1\cdot p_1 ~ q\cdot p_2 - k_1\cdot p_2 ~ q\cdot p_1)\big] \nn \\
&=& 2\big[ P\cdot p_1 ~ k_1\cdot s_{1x} ~ q\cdot p_2 - P\cdot p_2 ~ k_1\cdot s_{1x} ~ q\cdot p_1\big]  = 2 q^2 m^3 (\tau-1) \sqrt{\tau} \sin \theta, \nn
\ea
where $s_{1x}\cdot q=s_{1x}\cdot P=0$. These differences lead to
\ba
L_{\mu\nu}^{(v)}(S) H_{\mu\nu}^{(v)}(s_{1x}) &=& - 8 m^2 q^2 \sqrt{\tau} \sin \theta \big[|\tilde  G_M|^2 + Re (\tilde G_M(\tilde G_E - \tilde G_M)^*) \big] \nn \\
&=& - 8 m^2 q^2 \sqrt{\tau} \sin \theta Re (\tilde G_M \tilde G_E^*) \label{Px1term}
\ea
And step (\ref{step3})
\ba
L_{\mu\nu}^{(i)}(S) H_{\mu\nu}^{(i)}(s_{1x})&=&2 q^2 m A_{2\gamma} \big( (\tilde G_M + \tilde G_E)^*  s_{1x}\cdot p_2 + (\tilde G_M - \tilde G_E)^* s_{1x}\cdot p_1 \big) \nn \\
&&- 4 m A_{2\gamma} \big[ (\tilde G_M + \tilde G_E)^*  (k_2\cdot s_{1x} ~ k_1\cdot p_2 + k_1\cdot s_{1x} ~ k_2\cdot p_2) \nn \\
&&+ (\tilde G_M - \tilde G_E)^* (k_2\cdot s_{1x} ~ k_1\cdot p_1 + k_1\cdot s_{1x} ~ k_2\cdot p_1) \big] \nn \\
&=& 0 - 4 m A_{2\gamma} \big[ (\tilde G_M + \tilde G_E)^* k_1\cdot s_{1x} ~ p_2 \cdot (k_2-k_1)  \nn \\
&&+ (\tilde G_M - \tilde G_E)^* k_1\cdot s_{1x} ~ p_1\cdot (k_2-k_1) \big] \nn \\
&=& 16 m^4 \tau \sqrt{\tau-1} \cos \theta \sin \theta A_{2\gamma} \tilde G_E^* . \label{Px2term}
\ea
The results (\ref{Px1term}, \ref{Px2term}) lead to final formula for $P_x$
\ba
P_x = - \frac{2\sin \theta}{D \sqrt{\tau}} \left[ Re (\tilde G_M \tilde G_E^*) - \sqrt{\frac{\tau-1}{\tau}} \cos \theta Re (A_{2\gamma} \tilde G_E^*) \right].
\ea

\subsection{The component $P_z$}
$P_z$ is the polarization of the  outgoing antinucleon along the $z$-axis. It is a double spin polarization observable, induced by the polarization of incoming electron. The definition of $P_z$ is (similarly to $P_y$):
\be
P_z = \frac{1}{D q^4} \left[ L_{\mu\nu}^{(v)} H_{\mu\nu}^{(v)}(s_{1z}) +2 Re (L_{\mu\nu}^{(i)} H_{\mu\nu}^{(i)}(s_{1z}) ) \right], \nn
\ee
where $s_{1z}$ is the polarization four-vector with components (similar to longitudinal polarization of electron, Eq. (\ref{longpolel}) )
\ba
\vec{\xi}=(0,0,1) \Rightarrow s_{z0}=\frac{|\vec p_1|}{m}=\sqrt{\tau-1} \: ; \: \vec s_{1z}=(0,0,\frac{E}{m})=(0,0,\sqrt{\tau}).   \label{sz}
\ea
As we can see $s_{1z}$ doesn't have $y$-component, what implies (similar as for $P_x$) that the unpolarized electron part doesn't contribute to $P_z$
\ba
P_z = \frac{1}{D q^4} \left[ L_{\mu\nu}^{(v)}(S) H_{\mu\nu}^{(v)}(s_{1z}) +2 Re (L_{\mu\nu}^{(i)}(S) H_{\mu\nu}^{(i)}(s_{1z}) ) \right]. \label{Pzpred}
\ea 
The first part of Eq. (\ref{Pzpred}) comes from Eqs. (\ref{Lvmini1}, \ref{Hvmini1})
\ba
L_{\mu\nu}^{(v)}(S) H_{\mu\nu}^{(v)}(s_{1z}) &=& 2i \left\langle \mu \nu k_1 q \right\rangle \times \frac{2}{m(\tau-1)} \left[ i m^2 (\tau - 1) |\tilde G_M|^2 \left< \mu \nu q s_{1z} \right> \right . \nn \\
&&+ i Re(\tilde G_M (\tilde G_E-\tilde G_M)^*) ( P_{\mu} \left< \nu p_2 p_1 s_{1z} \right> - P_{\nu} \left< \mu p_2 p_1 s_{1z} \right> ) \nn \\
&& \left .+ Im (\tilde G_M \tilde G_E^*) (P_{\mu} \left< \nu p_2 p_1 s_{1z} \right> + P_{\nu} \left< \mu p_2 p_1 s_{1z} \right> ) \right ], \nn
\ea
where antisymmetric leptonic tensor gives vanish with symmetric parts of hadronic tensor
\ba
L_{\mu\nu}^{(v)}(S) H_{\mu\nu}^{(v)}(s_{1z}) &=& 2i \left\langle \mu \nu k_1 q \right\rangle \times \frac{2}{m(\tau-1)} \left[ i m^2 (\tau - 1) |\tilde G_M|^2 
\left< \mu \nu q s_{1z} \right> \right .\nn \\
&&\left .  + i Re(\tilde G_M (\tilde G_E-\tilde G_M)^*) ( P_{\mu} \left< \nu p_2 p_1 s_{1z} \right> - P_{\nu} \left< \mu p_2 p_1 s_{1z} \right> )\right  .], 
\label{pre1partPz}
\ea
where
\ba
\left\langle \mu \nu k_1 q \right\rangle \times \left< \mu \nu q s_{1z} \right> = 2 ( k_1\cdot s_{1z} ~ q^2 - q\cdot s_{1z} ~ k_1\cdot q)=- 2 m q^2 \tau \cos \theta \label{step1again}
\ea
and
\ba
&& \left\langle \mu \nu k_1 q \right\rangle \times ( P_{\mu} \left< \nu p_2 p_1 s_{1z} \right> - P_{\nu} \left< \mu p_2 p_1 s_{1z} \right> ) = -2 \left\langle \mu P k_1 q \right\rangle \times \left< \mu p_2 p_1 s_{1z} \right>  \nn \\
&& = -2 \big[ P\cdot p_2 ( k_1\cdot s_{1z} ~ q\cdot p_1 - k_1\cdot p_1 ~ q\cdot s_{1z}) + P\cdot p_1 (k_1\cdot p_2 ~ q\cdot s_{1z} - k_1\cdot s_{1z} ~ q\cdot p_2) \nn \\
&& ~~+ P\cdot s_{1z} (k_1\cdot p_1 ~ q\cdot p_2 - k_1\cdot p_2 ~ q\cdot p_1)\big] =0, \label{step2again}
\ea
where we used notations (\ref{CMSvectors}) and (\ref{sz}). Now we can use  Eqs. (\ref{step1again}, \ref{step2again}) in Eq. (\ref{pre1partPz})
\ba
L_{\mu\nu}^{(v)}(S) H_{\mu\nu}^{(v)}(s_{1z}) &=& 2 q^4 |\tilde G_M|^2. \cos \theta \label{Pz1part}
\ea
The second part of (\ref{Pzpred}) is according to (\ref{Limini1}, \ref{Himini1})
\ba
L_{\mu\nu}^{(i)}(S) H_{\mu\nu}^{(i)}(s_{1z}) &=& \left [q^2 g_{\mu\nu} - 2 (k_{2\mu} k_{1\nu} + k_{2\nu} k_{1\mu})\right ]  m A_{2\gamma} \left [ - 2q\cdot s_{1z} \tilde G_M^* g_{\mu\nu}  \right .\nn \\
&&-
\frac{2q\cdot s_{1z}}{m^2 (1-\tau)}(\tilde G_M-\tilde G_E)^*p_{1\mu} P_{\nu}  \nn \\
&&+ (\tilde G_M+\tilde G_E)^* (s_{1z\mu} p_{2\nu} + s_{1z\nu} p_{2\mu}) + (\tilde G_M - \tilde G_E)^* (s_{1z\mu} p_{1\nu} + s_{1z\nu} p_{1\mu}) \nn \\
&& \left .- (\tilde G_M + \tilde G_E)^* (s_{1z\mu} p_{1\nu} - s_{1z\nu} p_{1\mu}) - (\tilde G_M - \tilde G_E)^* (s_{1z\mu} p_{2\nu} - s_{1z\nu} p_{2\mu}) \right ], \nn
\ea
where the symmetric leptonic tensor vanish when multiplied with the antisymmetric part of hadronic tensor
\ba
L_{\mu\nu}^{(i)}(S) H_{\mu\nu}^{(i)}(s_{1z}) &=& \left [q^2 g_{\mu\nu} - 2 (k_{2\mu} k_{1\nu} + k_{2\nu} k_{1\mu})\right ]  \nn \\
&&\times  m A_{2\gamma} \left [ - 2q\cdot s_{1z} \tilde G_M^* g_{\mu\nu} - \frac{2q\cdot s_{1z}(\tilde G_M-\tilde G_E)^*}{m^2 (1-\tau)}p_{1\mu} P_{\nu}  \right .\nn \\
&&+ \left . (\tilde G_M+\tilde G_E)^* (s_{1z\mu} p_{2\nu} + s_{1z\nu} p_{2\mu}) + (\tilde G_M - \tilde G_E)^* (s_{1z\mu} p_{1\nu} + s_{1z\nu} p_{1\mu}) \right] \nn \\
\ea
and after multiplication we get
\ba
&&L_{\mu\nu}^{(i)}(S)H_{\mu\nu}^{(i)}(s_{1z})  = 2 q^2 m A_{2\gamma} \left [ - 3 q\cdot s_{1z} \tilde G_M^* -
\frac{q\cdot s_{1z}(\tilde G_M-\tilde G_E)^*}{m^2 (1-\tau)}p_1\cdot P + 2\tilde G_E^* s_{1z} \cdot P 
\right ] \nn \\
&& \hspace*{1truecm} +  4 m A_{2\gamma} \left [ 2q\cdot s_{1z}\tilde G_M^* k_1\cdot k_2 + 
\frac{q\cdot s_{1z}(\tilde G_M-\tilde G_E)^*} {m^2 (1-\tau)} 
\left (p_1\cdot k_1~P\cdot k_2 + p_1\cdot k_2 ~ P\cdot k_1\right ) \right .\nn \\
&& \hspace*{1truecm} - (\tilde G_M+\tilde G_E)^*(s_{1z}\cdot k_2 ~ p_2\cdot k_1 + s_{1z}\cdot k_1 ~ p_2\cdot k_2)  \nn \\
&& \hspace*{1truecm}- \left .
(\tilde G_M-\tilde G_E)^*(s_{1z}\cdot k_2 ~ p_1\cdot k_1 + s_{1z}\cdot k_1 ~ p_1\cdot k_2) \right ]. 
\label{eq:eqlk}
\ea
Inserting Eqs. (\ref{CMSvectors}) and (\ref{sz}) the following expression is obtained:
\be
L_{\mu\nu}^{(i)}(S) H_{\mu\nu}^{(i)}(s_{1z}) =-q^4 A_{2\gamma} \tilde G_M^* \sqrt{\frac{\tau-1}{\tau}} (1+\cos^2 \theta). \label{Pz2part}
\ee
Substituting Eqs. (\ref{Pz1part}, \ref{Pz2part}) into Eq. (\ref{Pzpred}),  the final formula for $P_z$ is:
\be
P_z = \frac{2}{D} \left [ |\tilde G_M|^2 \cos \theta - Re (A_{2\gamma} \tilde G_M^*) \sqrt{\frac{\tau-1}{\tau}} (1+ \cos^2 \theta) \right ].
 \label{eq:eqpz}
\ee 
Transversally polarized electron beams lead to
antinucleon polarization, which is a factor $(m_e/m)$ smaller than in case of 
longitudinal polarization.

The polarization component $P_z$ vanishes when the proton is emitted at an 
angle $\theta = 90^0$ in the Born approximation. But the presence of the TPE
term $A_{2\gamma}$ in the electromagnetic hadron current may lead to non--zero value 
of this quantity if the amplitude $A_{2\gamma}(\theta = 90^0)$ is not zero, since the 
value of this component is determined by the term $ReA_{2\gamma}G_M^*.$

\section{Spin correlations}

Let us consider the case when the produced antinucleon and nucleon
are both polarized. For convenience, let us write the vector part of the hadronic current in the following form
\be
J_{\mu }^{(v)}=\bar u(p_2)\left  [\tilde G_M\gamma_{\mu }
+G_2P_{\mu }\right  ]u(-p_1),
\label{eq:JMU}
\ee
where we introduce the following notation:
\be
G_2=-\frac{\tilde G_M-\tilde G_E}{m(1-\tau )},
\label{eq:eqf2}
\ee
which differs by sign for the corresponding Eqs. (\ref{hadronV},\ref{eq:eqfft}). 
Then, according to the definition, the hadronic tensor $H_{\mu\nu}^{(v)}$
is
\ba
H_{\mu\nu}^{(v)}&=&J_{\mu }^{(v)}J_{\nu }^{(v)*}=
\bar u(p_2)\left  [\tilde G_M\gamma_{\mu }+G_2P_{\mu }\right  ]u(-p_1)
\bar u(-p_1)\gamma_4\left  [\tilde G_M\gamma_{\nu }+
G_2P_{\nu }\right  ]^+\gamma_4u(p_2) \nn \\
&=&Tr\Lambda (p_2)\left  [\tilde G_M\gamma_{\mu }+G_2P_{\mu }\right  ]
\Lambda (-p_1)\left  [\tilde G_M^*\gamma_{\nu }+G_2^*P_{\nu }\right  ] \label{eq:eqi1} \\
&=&Tr\frac{1}{2}(\hat p_2+m)(1-\gamma_5\hat s_2)
\left  [\tilde G_M\gamma_{\mu }+G_2P_{\mu }\right  ]
\frac{1}{2}(\hat p_1-m)(1-\gamma_5\hat s_1)
\left  [\tilde G_M^*\gamma_{\nu }+G_2^*P_{\nu }\right  ], 
\nn
\ea
where $\Lambda (p_2)(\Lambda (-p_1))$ is the spin--density matrix of the
nucleon (antinucleon).

Retaining here the contribution proportional to the final particle
polarizations we obtain for the $H_{\mu\nu}^{(v)}(s_1, s_2)$ tensor the
following expression
\ba
H_{\mu\nu}^{(v)}(s_1, s_2)&=&\left(-\frac{1}{2}\right)\left(-\frac{1}{2}\right)
Tr(\hat p_2+m)\gamma_5\hat s_2\left  [\tilde G_M\gamma_{\mu }+
G_2P_{\mu }\right  ](\hat p_1-m)\gamma_5\hat s_1
\left  [\tilde G_M^*\gamma_{\nu }+G_2^*P_{\nu }\right  ] \nn\\
&=&\frac{1}{4}Tr(\hat p_2+m)\hat s_2\left  [-\tilde G_M\gamma_{\mu }+
G_2P_{\mu }\right  ](\hat p_1+m)\hat s_1\left  [\tilde G_M^*\gamma_{\nu }
+G_2^*P_{\nu }\right  ].
\label{eq:eqi3}
\ea
 $\gamma_5$ was eliminated using the following rules: $\gamma_5^2=1$ and
$\gamma_5\gamma_{\mu}+\gamma_{\mu}\gamma_5=0.$ So, the expression for this
tensor can be written as
\ba
H_{\mu\nu}^{(v)}(s_1, s_2)&=&\frac{1}{4}Tr\left  [G_2P_{\mu }(m\hat s_2+
\hat p_2\hat s_2)-m\tilde G_M\hat s_2\gamma_{\mu }-
\tilde G_M\hat p_2\hat s_2\gamma_{\mu }\right  ] \nn\\
&&\left  [G_2^*P_{\nu }(m\hat s_1+
\hat p_1\hat s_1)+m\tilde G_M^*\hat s_1\gamma_{\nu }+
\tilde G_M^*\hat p_1\hat s_1\gamma_{\nu }\right  ]. 
\label{eq:eqi4a}
\ea
Only the terms with even number of gamma matrices do not vanish:
\ba
&&H_{\mu\nu}^{(v)}(s_1, s_2)=\nn\\
&&~=
\frac{1}{4}
Tr\left[m^2|G_2|^2P_{\mu }P_{\nu }
\hat s_2\hat s_1-m\tilde G_MG_2^*P_{\nu }\hat p_2\hat s_2
\gamma_{\mu }\hat s_1
\right .
\nn\\
&&~~~
+|G_2|^2P_{\mu }P_{\nu }\hat p_2\hat s_2\hat p_1\hat s_1-
m\tilde G_MG_2^*P_{\nu }\hat s_2\gamma_{\mu }\hat p_1\hat s_1-
m^2|\tilde G_M|^2\hat s_2\gamma_{\mu }\hat s_1\gamma_{\nu}\nn\\
&&~~~
\left .
+m\tilde G_M^*G_2P_{\mu }\hat p_2\hat s_2\hat s_1\gamma_{\mu }+
m\tilde G_M^*G_2P_{\mu }\hat s_2\hat p_1\hat s_1\gamma_{\mu }-
|\tilde G_M|^2\hat p_2\hat s_2\gamma_{\mu }\hat p_1\hat s_1\gamma_{\nu}
\right ] \nn\\
&&~=m^2|G_2|^2P_{\mu }P_{\nu }s_1\cdot s_2+|G_2|^2P_{\mu }P_{\nu }
(p_1\cdot s_1p_2\cdot s_2+p_1\cdot s_2p_2\cdot s_1-
p_1\cdot p_2s_1\cdot s_2) \nn\\
&&~~~-m\tilde G_MG_2^*P_{\nu }(p_2\cdot s_2s_{1\mu}+p_2\cdot s_1s_{2\mu}-
s_1\cdot s_2p_{2\mu}+p_1\cdot s_1s_{2\mu}+s_1\cdot s_2p_{1\mu}-
p_1\cdot s_2s_{1\mu}) \nn\\
&&~~~-m^2|\tilde G_M|^2(s_{1\mu}s_{2\nu}+s_{2\mu}s_{1\mu}-
s_1\cdot s_2g_{\mu\nu})+m\tilde G_M^*G_2P_{\mu }(p_2\cdot s_2s_{1\nu} \nn\\
&&~~~-p_2\cdot s_1s_{2\nu}+s_1\cdot s_2p_{2\nu}+p_1\cdot s_1s_{2\nu}
-s_1\cdot s_2p_{1\nu}+p_1\cdot s_2s_{1\nu})-\frac{1}{4}|\tilde G_M|^2
X_{\mu\nu}, 
\label{eq:eqi5}
\ea
where $X_{\mu\nu}=Tr\hat p_2\hat s_2\gamma_{\mu }\hat p_1\hat s_1
\gamma_{\nu}$ (we calculate this trace separately). To simplify this
expression let us remind that the four--vectors $s_{i\mu} (i=1,2)$ satisfy to the conditions $p_1\cdot s_1=p_2\cdot s_2=0$. Applying  these relations we  obtain
\begin{subequations} \label{E:sp}
\begin{gather}
q\cdot s_1=(p_1+p_2)\cdot s_1=p_1\cdot s_1+p_2\cdot s_1=p_2\cdot s_1, \label{E:sp1}\\
q\cdot s_2=(p_1+p_2)\cdot s_2=p_1\cdot s_2+p_2\cdot s_2=p_1\cdot s_2. \label{E:sp2}
\end{gather}
\end{subequations}
Taking into account that
$$P_{\mu}=\frac{1}{2}(p_2-p_1)_{\mu},~ 
m^2-p_1\cdot p_2=2m^2(1-\tau ) $$
the expression for the $H_{\mu\nu}^{(v)}(s_1, s_2)$ tensor can be written in the
following form
\ba
H_{\mu\nu}^{(v)}(s_1, s_2)&=&|G_2|^2P_{\mu }P_{\nu }[q\cdot s_1q\cdot s_2+
2m^2(1-\tau )s_1\cdot s_2]\nn\\
&&-m^2|\tilde G_M|^2(s_{1\mu}s_{2\nu}+
s_{1\nu}s_{2\mu}-s_1\cdot s_2g_{\mu\nu}) \nn\\
&&
-m\tilde G_MG_2^*P_{\nu }(q\cdot s_1s_{2\mu}-q\cdot s_2s_{1\mu}-
2s_1\cdot s_2P_{\mu}) \nn\\
&&
+m\tilde G_M^*G_2P_{\mu }(q\cdot s_2s_{1\nu}-q\cdot s_1s_{2\nu}+
2s_1\cdot s_2P_{\nu})-\frac{1}{4}|\tilde G_M|^2X_{\mu\nu}. 
\label{eq:eqi6}
\ea
Let us calculate the $X_{\mu\nu}$ tensor, which can be written in the following form
$$X_{\mu\nu}=Tr\hat p_2\hat s_2\gamma_{\mu }\hat p_1\hat s_1\gamma_{\nu}=
Tr(\gamma_{i}\gamma_{k}\gamma_{\mu}\gamma_{m}\gamma_{n}\gamma_{\nu })
p_{2i}s_{2k}p_{1m}s_{1n}.$$
The trace of six gamma matrices can be written as
$$Tr\gamma_{i}\gamma_{k}\gamma_{\mu}\gamma_{m}\gamma_{n}\gamma_{\nu }=
g_{ik}\{\mu mn\nu\}-g_{i\mu}\{kmn\nu\}+g_{im}\{k\mu n\nu\}
-g_{in}\{k\mu m\nu\}+g_{i\nu}\{k\mu mn\}, $$
where we introduce the notation $\{iklm\}=4(g_{ik}g_{lm}-g_{il}g_{km}+
g_{im}g_{kl}).$ So, the $X_{\mu\nu}$ tensor becomes:
\ba
X_{\mu\nu}&=&p_2\cdot s_2\{\mu mn\nu\}p_{1m}s_{1n}-
\{kmn\nu\}p_{2\mu}s_{2k}p_{1m}s_{1n}+p_1\cdot p_2\{k\mu n\nu\}s_{2k}s_{1n} \nn\\
&&-s_1\cdot p_2\{k\mu m\nu\}s_{2k}p_{1m}+\{k\mu mn\}p_{2\nu}s_{2k}p_{1m}s_{1n} \nn\\
&=&-p_{2\mu}4(p_1\cdot s_2s_{1\nu}+p_1\cdot s_1s_{2\nu}-s_1\cdot s_2p_{1\nu})
+p_1\cdot p_24(s_{1\mu}s_{2\nu}+s_{1\nu}s_{2\mu}-s_1\cdot s_2g_{\mu\nu})\nn\\
&&-s_1\cdot p_24(s_{2\mu}p_{1\nu}+s_{2\nu}p_{1\mu}-p_1\cdot s_2g_{\mu\nu})+
p_{2\nu}4(s_{2\mu}s_1\cdot p_1+p_{1\mu}s_1\cdot s_2-s_{1\mu}p_1\cdot s_2) \nn \\
&=&
4\left  [p_1\cdot p_2(s_{1\mu}s_{2\nu}+s_{1\nu}s_{2\mu}-
s_1\cdot s_2g_{\mu\nu})-q\cdot s_2(s_{1\mu}p_{2\nu}+s_{1\nu}p_{2\mu})+
q\cdot s_1q\cdot s_2g_{\mu\nu} \right .\nn \\
&&\left  . -q\cdot s_1(s_{2\mu}p_{1\nu}+s_{2\nu}p_{1\mu})+
s_1\cdot s_2(p_{1\mu}p_{2\nu}+p_{1\nu}p_{2\mu})\right ].
\label{eq:eqi6a}
\ea 
So, using the relations $p_{1\nu}=q_{\nu}/2-P_{\nu},$ $p_{2\nu}=q_{\nu}/2
+P_{\nu},$ we have for the $H_{\mu\nu}^{(v)}(s_1, s_2)$ tensor\footnote{We omitted the
terms proportional to $q_{\mu}$ or $q_{\nu}$, since they do not contribute
to the cross section and to the polarization observables due to the
conservation of the leptonic current.}
\ba
H_{\mu\nu}^{(v)}(s_1, s_2)&=&|G_2|^2P_{\mu }P_{\nu }\left  [
q\cdot s_1q\cdot s_2+2m^2(1-\tau )s_1\cdot s_2\right ]
- m^2|\tilde G_M|^2(s_{1\mu}s_{2\nu}+s_{1\nu}s_{2\mu}  \nn\\
&&-s_1\cdot s_2g_{\mu\nu})+4ms_1\cdot s_2ReG_2\tilde G_M^*P_{\mu}P_{\nu}
+mq\cdot s_2(G_2\tilde G_M^*P_{\mu}s_{1\nu}+
G_2^*\tilde G_MP_{\nu}s_{1\mu})  \nn\\
&&-mq\cdot s_1(G_2\tilde G_M^*P_{\mu}s_{2\nu}+
G_2^*\tilde G_MP_{\nu}s_{2\mu})-|\tilde G_M|^2\left  [
p_1\cdot p_2(s_{1\mu}s_{2\nu}+s_{1\nu}s_{2\mu} \right .\nn\\
&&-s_1\cdot s_2g_{\mu\nu})-q\cdot s_2(s_{1\nu}P_{\mu}+
s_{1\mu}P_{\nu})+q\cdot s_1(s_{2\nu}P_{\mu}+s_{2\mu}P_{\nu}) \nn\\
&&+\left . q\cdot s_1q\cdot s_2g_{\mu\nu}-2s_1\cdot s_2P_{\mu}P_{\nu}\right  ]. 
\label{eq:eqi7}
\ea
Using the relation $\psi a_ib_j+\psi^* a_jb_i=Re\psi (a_ib_j+a_jb_i)+
iIm\psi (a_ib_j-a_jb_i)$ we can write Eq.~(\ref{eq:eqi7}) in the
following general form
\ba
H_{\mu\nu}^{(v)}(s_1, s_2)&=&C_1g_{\mu\nu }+C_2P_{\mu }P_{\nu }+
C_3(P_{\mu }s_{1\nu }+P_{\nu }s_{1\mu})+C_4(P_{\mu }s_{2\nu }+
P_{\nu }s_{2\mu})\nn\\
&&
+C_5(s_{1\mu }s_{2\nu }+s_{1\nu }s_{2\mu})+
iC_6(P_{\mu }s_{1\nu }-P_{\nu }s_{1\mu})+
iC_7(P_{\mu }s_{2\nu }-P_{\nu }s_{2\mu}). 
\label{eq:eqi8}
\ea
The structure functions $C_i$ have the following form
\ba
C_1&=&\frac{1}{2}(q^2s_1\cdot s_2-2q\cdot s_1q\cdot s_2)|\tilde G_M|^2, \nn\\
C_2&=&\frac{2}{(\tau -1)^2}\left  [|\tau\tilde G_M-\tilde G_E|^2
s_1\cdot s_2+\frac{1}{4m^2}(2q\cdot s_1q\cdot s_2-q^2s_1\cdot s_2)
|\tilde G_E-\tilde G_M|^2\right  ], \nn\\
C_3&=&ReE_1,~ C_4=ReE_2,~
C_5=-\frac{q^2}{2}|\tilde G_M|^2,~C_6=ImE_1, ~ C_7=ImE_2, \nn\\
E_1&=&\frac{q\cdot s_2}{\tau -1}(\tau |\tilde G_M|^2-
\tilde G_E\tilde G_M^*), \ \
E_2=-\frac{q\cdot s_1}{\tau -1}(\tau |\tilde G_M|^2-
\tilde G_E\tilde G_M^*), \label{eq:eqi4}
\ea

Now let us calculate the hadronic tensor $H_{\mu\nu}^{(i)}(s_1, s_2)$. According to
the definition one has
\ba
H_{\mu\nu}^{(i)}(s_1, s_2)&=&J_{\mu }^{(a)}J_{\nu }^{(v)*}
=\bar u(p_2)\gamma_{\mu}\gamma_5u(-p_1)A_{2\gamma}
\bar u(-p_1)\left  [\tilde G_M^*\gamma_{\nu }+
G_2^*P_{\nu }\right  ]u(p_2) \label{eq:eqi9}\\
&=&
A_{2\gamma} Tr\Lambda (p_2)\gamma_{\mu}\gamma_5\Lambda (-p_1)
\left [\tilde G_M^*\gamma_{\nu }+G_2^*P_{\nu }\right ] \nn\\
&=&A_{2\gamma}Tr\frac{1}{2}(\hat p_2+m)(1-\gamma_5\hat s_2)
\gamma_{\mu}\gamma_5\frac{1}{2}(\hat p_1-m)(1-\gamma_5\hat s_1)
\left [\tilde G_M^*\gamma_{\nu }+G_2^*P_{\nu }\right ]. 
\nn
\ea
Retaining here the contribution proportional to the final particle
polarizations we obtain for the $H_{\mu\nu}^{(i)}(s_1, s_2)$ tensor the
following expression
\ba
H_{\mu\nu}^{(i)}(s_1, s_2)&=&(-\frac{1}{2})(-\frac{1}{2})A_{2\gamma}
Tr(\hat p_2+m)\gamma_5\hat s_2\gamma_{\mu}\gamma_5
(\hat p_1-m)\gamma_5\hat s_1\left [\tilde G_M^*\gamma_{\nu }
+G_2^*P_{\nu }\right ]\nn \\
&=&\frac{1}{4}A_{2\gamma}Tr(\hat p_2+m)\hat s_2\gamma_{\mu}
(\hat p_1-m)\gamma_5\hat s_1\left [\tilde G_M^*\gamma_{\nu }
+G_2^*P_{\nu }\right ], \label{eq:eqi10}
\ea
where the following properties were applied: $\gamma_5\hat s_2\gamma_{\mu}\gamma_{5}=$
$-\hat s_2\gamma_5\gamma_{\mu}\gamma_{5}=$ $\hat s_2\gamma_{\mu}
\gamma_{5}^2=$ $\hat s_2\gamma_{\mu}$ and $TrABC=TrBCA$. Therefore:
\ba
H_{\mu\nu}^{(i)}(s_1, s_2)&=&\frac{1}{4}A_{2\gamma}Tr\gamma_5\left  [
\tilde G_M^*\hat s_1\gamma_{\nu}+G_2^*P_{\nu }\hat s_1\right  ]
\left  [-m^2\hat s_2\gamma_{\mu}-m\hat p_2\hat s_2\gamma_{\mu}+
m\hat s_2\gamma_{\mu}\hat p_1+\hat p_2\hat s_2\gamma_{\mu}\hat p_1\right  ]\nn \\
&=&\frac{1}{4}A_{2\gamma}Tr\gamma_5\left  [mG_2^*P_{\nu}(\hat s_1\hat s_2
\gamma_{\mu}\hat p_1-\hat s_1\hat p_2\hat s_2\gamma_{\mu})-
m^2\tilde G_M^*\hat s_1\gamma_{\nu}\hat s_2\gamma_{\mu}+
\tilde G_M^*\hat s_1\gamma_{\nu}
\hat p_2\hat s_2\gamma_{\mu}\hat p_1\right  ] \nn \\
&=&\frac{1}{4}A_{2\gamma}(-4i)\left  [mG_2^*P_{\nu}(<s_1s_2\mu p_1>-
<s_1p_2s_2\mu >)-m^2\tilde G_M^*<s_1\nu s_2\mu >\right  ]
\nn \\
&&
+\frac{1}{4}A_{2\gamma}\tilde G_M^*Y_{\mu\nu}, 
\label{eq:eql}
\ea
where $Y_{\mu\nu}=Tr\gamma_5\hat s_1\gamma_{\nu}
\hat p_2\hat s_2\gamma_{\mu}\hat p_1.$ In order to calculate this trace
let us write the following identity:
$$\gamma_{\alpha}\gamma_{\beta}\gamma_{\mu}=g_{\alpha\beta}\gamma_{\mu}+
g_{\beta\mu}\gamma_{\alpha}-g_{\alpha\mu}\gamma_{\beta}
+i<\alpha\beta\mu\nu >\gamma_{\nu}\gamma_{5}. $$
So, one finds:
\ba
Y_{\mu\nu}&=&Tr\gamma_5\hat s_1\gamma_{\nu}\hat p_2\hat s_2
\gamma_{\mu}\hat p_1=Tr\gamma_5\gamma_{\alpha}\gamma_{\nu}
\gamma_{\beta}\hat s_2\gamma_{\mu}\hat p_1s_{1\alpha}p_{2\beta}  \nn \\
&=&Tr\gamma_5\left  [g_{\alpha\nu}\gamma_{\beta}+
g_{\beta\nu}\gamma_{\alpha}-g_{\alpha\beta}\gamma_{\nu}
+i<\alpha\nu\beta\rho >\gamma_{\rho}\gamma_{5}\right  ]
\hat s_2\gamma_{\mu}\hat p_1s_{1\alpha}p_{2\beta}\nn \\
&=& Tr\gamma_5\left  [s_{1\nu}\hat p_2+p_{2\nu}\hat s_1-
p_2\cdot s_1\gamma_{\nu}+i<s_1\nu p_2\rho >\gamma_{\rho}\gamma_{5}\right  ]
\hat s_2\gamma_{\mu}\hat p_1\nn \\
&=&-4i(s_{1\nu}<p_2s_2\mu p_1>+p_{2\nu}<s_1s_2\mu p_1>-
p_2\cdot s_1<\nu s_2\mu p_1>) \nn\\
&&
-4i<s_1\nu p_2\rho >(s_{2\rho}p_{1\mu}+
s_{2\mu}p_{1\rho}-p_1\cdot s_2g_{\rho\mu})\nn \\
&=&-4i(s_{1\nu}<p_2s_2\mu p_1>+p_{2\nu}<s_1s_2\mu p_1>-
p_2\cdot s_1<\nu s_2\mu p_1>+
s_{2\mu}<s_1\nu p_2p_1>\nn \\
&&+p_{1\mu}<s_1\nu p_2s_2>-
p_1\cdot s_2<s_1\nu p_2\mu >). \nn
\label{eq:eqi9a}
\ea
Then, the expression for the $H_{\mu\nu}^{(i)}(s_1, s_2)$ tensor can
be written as 
\ba
H_{\mu\nu}^{(i)}(s_1, s_2)&=&-iA_{2\gamma}\left  [-2mG_2^*P_{\nu}<\mu s_1s_2P>-
m^2\tilde G_M^*<\mu\nu s_1s_2> \right .\nn \\
&&
 +\tilde G_M^*(-q\cdot s_1
<\mu\nu s_2p_1>-q\cdot s_2<\mu\nu s_1p_2>+
p_{1\mu}<\nu s_1s_2p_2>
\nn \\
&&
\left .+p_{2\nu}<\mu s_1s_2p_1>+s_{2\mu}<\nu p_2s_1p_1>+s_{1\nu}<\mu p_2s_2p_1>)\right  ]\nn  \\
&=&iA_{2\gamma}\left  [\frac{2}{\tau -1}(G_M-
G_E)^*P_{\nu}<\mu s_1s_2P>+G_M^*(m^2<\mu\nu s_1s_2> \right .\nn  \\
&&+q\cdot s_1<\mu\nu s_2p_1>+q\cdot s_2<\mu\nu s_1p_2>-
p_{1\mu}<\nu s_1s_2p_2> \nn \\
&&\left .-p_{2\nu}<\mu s_1s_2p_1>-s_{2\mu}<\nu p_2s_1p_1>-
s_{1\nu}<\mu p_2s_2p_1>)\right ]. 
\label{eq:eqi10a}
\ea
This tensor does not contain the terms proportional to $\Delta G_M$
and $\Delta G_E$ since it is proportional to the TPE term $A_{2\gamma}$
(terms of the order of $\alpha^2 $ compared
to the dominant (Born) terms were neglected). The antisymmetrical  parts of the $H_{\mu\nu}^{(v)}(s_1, s_2)$ and
$H_{\mu\nu}^{(i)}(s_1, s_2)$ tensors (with respect to the
$\mu $ and $\nu $ indices) arise due to the fact that nucleon
FFs in the TL region of the momentum transfer are complex quantities.

\subsection{Spin correlations: unpolarized electron beam }
Let us calculate the components of the polarization correlation tensor
$P_{ik}, (i,k=x,y,z),$ of the antinucleon and nucleon in presence of TPE
mechanism. Let us consider firstly the case of unpolarized lepton beams. The contribution of the vector part of the hadronic current is
determined by the following expression
\ba
S^{(v)}(s_1, s_2)&=&L_{\mu\nu}^{(v)}(0)H_{\mu\nu}^{(v)}(s_1, s_2)\nn  \\
&=&
\left  [-q^2g_{\mu\nu}+2(k_{1\mu}k_{2\nu}+k_{1\nu}k_{2\mu})\right  ]
\left  [C_1g_{\mu\nu}+C_2P_{\mu}P_{\nu}+C_3(P_{\mu}s_{1\nu}+
P_{\nu}s_{1\mu}) \right .\nn  \\
&&
+C_4(P_{\mu}s_{2\nu}+P_{\nu}s_{2\mu})+C_5(s_{1\mu}s_{2\nu}+
s_{1\nu}s_{2\mu})+iC_6(P_{\mu}s_{1\nu}-P_{\nu}s_{1\mu})
\nn  \\
&&
\left .
+
iC_7(P_{\mu}s_{2\nu}-P_{\nu}s_{2\mu})\right  ]\nn  \\
&=&-2q^2C_1+(4P\cdot k_1P\cdot k_2-q^2P^2)C_2+2\left  [
2(P\cdot k_1s_1\cdot k_2+P\cdot k_2s_1\cdot k_1)\right .
\nn  \\
&&
\left .
-q^2P\cdot s_1 \right  ]C_3 
+2\left  [2(P\cdot k_1s_2\cdot k_2+P\cdot k_2s_2\cdot k_1)-q^2P\cdot s_2
\right  ]C_4\nn  \\
&&
+2\left  [2(s_1\cdot k_1s_2\cdot k_2+s_2\cdot k_1s_1\cdot k_2)
-q^2s_1\cdot s_2\right  ]C_5. 
\label{eq:eqp1}
\ea
Note that the convolution of the symmetrical lepton tensor and antisymmetrical parts of the hadronic tensor (with respect to $\mu $ and $\nu $), which is 
proportional to the structure functions $C_6$ and $C_7$, is equal to zero.
Let us give some necessary relations among the kinematical variables:
\ba
&&P\cdot s_1=\frac{1}{2}(p_2-p_1)\cdot s_1=\frac{1}{2}p_2\cdot s_1=
\frac{1}{2}q\cdot s_1, \nn \\
&&
P\cdot s_2=\frac{1}{2}(p_2-p_1)\cdot s_2=-\frac{1}{2}p_1\cdot s_2=
-\frac{1}{2}q\cdot s_2, \label{eq:eqp2a} \\
&&P^2=\frac{1}{4}(2m^2-2p_1\cdot p_2)=\frac{1}{4}\left  [2m^2-
(q^2-2m^2)\right  ]=m^2(1-\tau ), \nn \\
&&P\cdot k_2+P\cdot k_1=P\cdot (k_1+k_2)=P\cdot q=
\frac{1}{2}(p_2-p_1)\cdot (p_2+p_1)=0\to P\cdot k_2=-P\cdot k_1
\nn
\ea
With the help of relations (\ref{eq:eqp2a}),  Eq. (\ref{eq:eqp1}) can be
 simplified,as:
\ba
S^{(v)}(s_1, s_2)&=&-2q^2C_1-\left  [4(P\cdot k_1)^2+(1-\tau )m^2q^2
\right  ]C_2+\left  [4P\cdot k_1s_1\cdot (k_2-k_1)-q^2q\cdot s_1
\right  ]C_3 \nn \\
&&
+\left  [4P\cdot k_1s_2\cdot (k_2-k_1)+q^2q\cdot s_2\right  ]C_4
\nn \\
&&
+2\left  [2(s_1\cdot k_1s_2\cdot k_2+s_1\cdot k_2s_2\cdot k_1)
-q^2s_1\cdot s_2\right  ]C_5. \label{eq:eqp2}
\ea
Substituting the expressions (\ref{eq:eqi4}) for the structure functions $C_i$ into Eq. (\ref{eq:eqp2}) one finds:
\ba
S^{(v)}(s_1, s_2)&=&-\frac{2}{\tau -1}\left  [\tau |\tilde G_M|^2-
|\tilde G_E|^2\right  ]\left  [4(P\cdot k_1)^2+(1-\tau )m^2q^2
\right  ]s_1\cdot s_2\nn \\
&&+\left  \{\frac{2q^2}{\tau -1}\left  [Re\tilde G_M\tilde G_E^*-
|\tilde G_M|^2\right  ]-\left  [4(P\cdot k_1)^2+(1-\tau )m^2q^2
\right  ]|G_2|^2\right  \}q\cdot s_1q\cdot s_2\nn \\
&&+4\frac{P\cdot k_1}{\tau -1}\left  [\tau |\tilde G_M|^2-
Re\tilde G_M\tilde G_E^*\right  ]\left  [
s_1\cdot (k_2-k_1)q\cdot s_2-s_2\cdot (k_2-k_1)q\cdot s_1\right  ] \nn \\
&&-2q^2|\tilde G_M|^2(s_1\cdot k_1s_2\cdot k_2+
s_1\cdot k_2s_2\cdot k_1). \label{eq:eqp4}
\ea
The scalar products of various four--vectors in the chosen
coordinate system is:
$$P\cdot k_1={\vec k}\cdot {\vec p}=Ep\cos\theta ,~
4(P\cdot k_1)^2+(1-\tau )m^2q^2=-p^2q^2\sin^2\theta ,~
\tau -1=\frac{p^2}{m^2}, $$
$$q\cdot s_1=2Es_{10}, \ q\cdot s_2=2Es_{20}, \
(k_2-k_1)\cdot s_1=2{\vec k}\cdot {\vec s}_1, \
(k_2-k_1)\cdot s_2=2{\vec k}\cdot {\vec s}_2. $$
The time and space components of the antinucleon (nucleon) polarization
four--vectors $s_{1\mu} (s_{2\mu})$ can be related to the unit
polarization vector ${\vec \xi}_1 ({\vec \xi}_2)$ in its rest frame. 
The following relations hold:
$$s_{10}=\frac{1}{m}{\vec p}\cdot {\vec \xi}_1=\frac{p}{m}\xi_{1z},~
{\vec s}_1={\vec \xi}_1+\frac{{\vec p}\cdot {\vec \xi}_1{\vec p}}{m(E+m)}, \
s_{1x}=\xi_{1x}, \ s_{1y}=\xi_{1y}, \ s_{1z}=\frac{E}{m}\xi_{1z}, $$
$$s_{20}=-\frac{1}{m}{\vec p}\cdot {\vec \xi}_2=-\frac{p}{m}\xi_{2z}, \
{\vec s}_2={\vec \xi}_2+\frac{{\vec p}\cdot {\vec \xi}_2{\vec p}}{m(E+m)}, \
s_{2x}=\xi_{2x}, \ s_{2y}=\xi_{2y}, \ s_{2z}=\frac{E}{m}\xi_{2z}. $$
Then the different scalar products, including polarization four--vectors,
are 
$$s_1\cdot s_2=-\xi_{1x}\xi_{2x}-\xi_{1y}\xi_{2y}-
\frac{E^2+p^2}{m^2}\xi_{1z}\xi_{2z}, \ \
q\cdot s_1q\cdot s_2=-\frac{p^2}{m^2}q^2\xi_{1z}\xi_{2z}, $$
\ba 
&&(k_2-k_1)\cdot s_1q\cdot s_2-(k_2-k_1)\cdot s_2q\cdot s_1=\nn \\
&&
= 
-\frac{p}{m}q^2\left  [\left (\frac{E}{m}\cos\theta\xi_{2z}-
\sin\theta\xi_{2x}\right)\xi_{1z}\right .\left .
+\left (\frac{E}{m}\cos\theta\xi_{1z}-
\sin\theta\xi_{1x}\right)\xi_{2z}\right  ], \nn \\
&&(s_1\cdot k_1s_2\cdot k_2+s_1\cdot k_2s_2\cdot k_1)=\nn \\
&&
=
-\frac{q^2}{2}\left  [\sin^2\theta\xi_{1x}\xi_{2x}-
\cos\theta \sin\theta \frac{E}{m}\left (\xi_{1x}\xi_{2z}+\xi_{2x}\xi_{1z}\right )+
\frac{1}{m^2}\left(p^2+E^2\cos^2\theta \right)\xi_{1z}\xi_{2z}\right  ]. 
\nn \ea
Eq. (\ref{eq:eqp4}) can be expressed in terms of the polarization
unit vectors ${\vec \xi}_1$ and ${\vec \xi}_2$ as:
\ba
S^{(v)}(s_1, s_2)&=&2m^2q^2\sin^2\theta \left  [\tau |\tilde G_M|^2-
|\tilde G_E|^2\right  ]s_1\cdot s_2\nn \\
&&
+
\frac{q^2}{p^2}\left  \{
2m^2\left  [Re\tilde G_E\tilde G_M^*-|\tilde G_M|^2\right  ]
+p^4\sin^2\theta |G_{N}|^2\right  \}q\cdot s_1q\cdot s_2\nn \\
&&
+4m^2\frac{E}{p}\cos\theta \left  [\tau |\tilde G_M|^2-
Re\tilde G_E\tilde G_M^*\right  ]
\left  [(k_2-k_1)\cdot s_1q\cdot s_2-(k_2-k_1)\cdot s_2q\cdot s_1\right  ] \nn \\
&&
-2q^2|\tilde G_M|^2(s_1\cdot k_1s_2\cdot k_2+
s_1\cdot k_2s_2\cdot k_1) \nn \\
&=&
-2m^2q^2\sin^2\theta \left  [\tau |\tilde G_M|^2-
|\tilde G_E|^2\right  ](\xi_{1x}\xi_{2x}+\xi_{1y}\xi_{2y}+
\frac{E^2+p^2}{m^2}\xi_{1z}\xi_{2z})\nn \\
&&-\frac{q^4}{m^2}\left  \{
2m^2\left  [Re\tilde G_E\tilde G_M^*-|\tilde G_M|^2\right  ]+
p^4\sin^2\theta |G_{2}|^2\right  \}\xi_{1z}\xi_{2z}\nn \\
&&-4mEq^2\cos\theta \left  [\tau |\tilde G_M|^2-
Re\tilde G_E\tilde G_M^*\right  ]\left  [
\left (\frac{E}{m}\cos\theta\xi_{2z}-\sin\theta\xi_{2x}\right )\xi_{1z}\right .\nn \\
&&+\left . \left(\frac{E}{m}\cos\theta\xi_{1z}-\sin\theta\xi_{1x}\right )\xi_{2z}\right  ]
\nn \\
&&
+q^4|\tilde G_M|^2\left  [\sin^2\theta\xi_{1x}\xi_{2x}-
\cos\theta \sin\theta \frac{E}{m}
\left (\xi_{1x}\xi_{2z} +\xi_{2x}\xi_{1z})+\frac{1}{m^2}(p^2+E^2\cos^2\theta \right )
\xi_{1z}\xi_{2z}\right  ] \nn \\
&=&A_1\xi_{1x}\xi_{2x}+A_2\xi_{1y}\xi_{2y}+A_3\xi_{1z}\xi_{2z}+
A_4\xi_{1x}\xi_{2z}+A_5\xi_{2x}\xi_{1z}, 
\label{eq:eqp5}
\ea
where the coefficients $A_i, (i=1-5)$, are
\ba
A_1&=&\frac{1}{2\tau}q^4\sin^2\theta\left  [|\tilde G_E|^2+
\tau |\tilde G_M|^2\right  ], \nn \\
A_2&=&\frac{1}{2\tau}q^4\sin^2\theta\left  [|\tilde G_E|^2-
\tau |\tilde G_M|^2\right  ], \nn \\
A_3&=&\frac{1}{2\tau}q^4\sin^2\theta\left  [(1+\cos^2\theta )
\tau |\tilde G_M|^2-|\tilde G_E|^2\right  ], \nn \\
A_4&=&A_5=-\frac{1}{\sqrt{\tau}}q^4\sin\theta \cos\theta
Re\tilde G_E\tilde G_M^*. 
\label{eq:eqp6}
\ea
Now let us consider the contribution of the axial part of the hadronic
current. The contraction of the unpolarized interference lepton tensor
and interference hadronic tensor is 
\ba
S^{(i)}(s_1, s_2)&=&L_{\mu\nu}^{(i)}(0)H_{\mu\nu}^{(i)}(s_1, s_2)\nn \\
&=&-2i<\mu\nu k_1k_2>
iA_{2\gamma}\left[\frac{2}{\tau -1}(G_M-G_E)^*P_{\nu}<\mu s_1s_2P> \right . \nn \\
&&+G_M^*(m^2<\mu\nu s_1s_2>+
q\cdot s_1<\mu\nu s_2p_1>+q\cdot s_2<\mu\nu s_1p_2>\nn \\
&&\left .-
p_{1\mu}<\nu s_1s_2p_2>-p_{2\nu}<\mu s_1s_2p_1>-s_{2\mu}<\nu p_2s_1p_1>-
s_{1\nu}<\mu p_2s_2p_1>)\right ] \nn \\
&=&2A_{2\gamma}\left [\frac{2}{\tau -1}(G_M-G_E)^*B_1+
G_M^*\left (m^2B_2+q\cdot s_1B_3+q\cdot s_2B_4\right .\right .
\nn \\
&&
 \left .+B_5+B_6+B_7+B_8\right) \Big ], 
\label{eq:eqp7}
\ea
where we introduce the notations for the contractions of the tensors
\ba
B_1&=&<\mu Pk_1k_2><\mu s_1s_2P>, ~B_2=<\mu\nu k_1k_2><\mu\nu s_1s_2>, \nn \\
B_3&=&<\mu\nu k_1k_2><\mu\nu s_2p_1>, ~ 
B_4=<\mu\nu k_1k_2><\mu\nu s_1p_2>,
\nn \\
B_5&=&<\mu p_1k_1k_2><\mu s_1s_2p_2>,~
B_6=-<\mu p_2k_1k_2><\mu s_1s_2p_1>,\nn \\ 
B_7&=&<\mu s_2k_1k_2><\mu p_2s_1p_1>, ~
B_8=-<\mu s_1k_1k_2><\mu p_2s_2p_1>. 
\label{eq:eqp8c}
\ea
Moreover
\ba
B_5+B_6&=&B_5-<\mu p_2k_1k_2><\mu s_1s_2(q-p_2)>
\nn \\
&=&
B_5-<\mu p_2k_1k_2>(<\mu s_1s_2q>-<\mu s_1s_2p_2>)  \nn \\
&=&<\mu p_1k_1k_2><\mu s_1s_2p_2>+<\mu p_2k_1k_2><\mu s_1s_2p_2>-
<\mu p_2k_1k_2><\mu s_1s_2q> \nn \\
&=&<\mu s_1s_2p_2><\mu (p_1+p_2)k_1k_2>-
<\mu p_2k_1k_2><\mu s_1s_2q>
\nn \\
&=&
-<\mu p_2k_1k_2><\mu s_1s_2q>. \label{eq:eqp9}
\ea 
The following
relations are used to calculate the quantities $B_i$:
\ba
<\mu\nu ab><\mu\nu cd>&=&2(a\cdot db\cdot c-a\cdot cb\cdot d), \nn \\
<\mu abc><\mu def>&=&a\cdot d(b\cdot fc\cdot e-b\cdot ec\cdot f)+
a\cdot e(b\cdot dc\cdot f-b\cdot fc\cdot d) \nn \\
&&+a\cdot f(b\cdot ec\cdot d-b\cdot dc\cdot e), \label{eq:eqp9a}
\ea 
which results in
\ba
B_1&=&P\cdot s_1(P\cdot k_1k_2\cdot s_2-P\cdot k_2k_1\cdot s_2)+
P\cdot s_2(P\cdot k_2k_1\cdot s_1-P\cdot k_1k_2\cdot s_1)\nn \\
&&
+P^2(k_1\cdot s_2k_2\cdot s_1-k_1\cdot s_1k_2\cdot s_2), \nn \\
B_2&=&2(k_1\cdot s_2k_2\cdot s_1-k_1\cdot s_1k_2\cdot s_2), \
B_3=2(k_1\cdot p_1k_2\cdot s_2-k_1\cdot s_2k_2\cdot p_1),\nn \\ 
B_4&=&2(k_1\cdot p_2k_2\cdot s_1-k_1\cdot s_1k_2\cdot p_2), \nn \\
B_5&+&B_6=-p_2\cdot s_1(q\cdot k_1k_2\cdot s_2-q\cdot k_2k_1\cdot s_2)-
p_2\cdot s_2(q\cdot k_2k_1\cdot s_1-q\cdot k_1k_2\cdot s_1)\nn \\
&&
-q\cdot p_2(k_1\cdot s_2k_2\cdot s_1-k_1\cdot s_1k_2\cdot s_2), \nn \\
B_7&=&p_2\cdot s_2(p_1\cdot k_1k_2\cdot s_1-p_1\cdot k_2k_1\cdot s_1)+
s_1\cdot s_2(k_1\cdot p_2k_2\cdot p_1-p_1\cdot k_1k_2\cdot p_2) \nn \\
&&+p_1\cdot s_2(k_1\cdot s_1k_2\cdot p_2-k_1\cdot p_2k_2\cdot s_1), \nn \\
B_8&=&-p_2\cdot s_1(p_1\cdot k_1k_2\cdot s_2-p_1\cdot k_2k_1\cdot s_2)-
s_1\cdot s_2(k_1\cdot p_2k_2\cdot p_1-p_1\cdot k_1k_2\cdot p_2)\nn \\
&& -p_1\cdot s_1(k_1\cdot s_2k_2\cdot p_2-k_1\cdot p_2k_2\cdot s_2)
\label{eq:eqp8a}
\ea 
which can be further simplified:
\ba
&&B_1=P\cdot k_1q\cdot s_1q\cdot s_2+m^2(1-\tau )(k_1\cdot s_2
k_2\cdot s_1-k_1\cdot s_1k_2\cdot s_2), \nn \\
&&B_5+B_6=-\frac{q^2}{2}q\cdot s_1(k_2-k_1)\cdot s_2-
\frac{q^2}{2}(k_1\cdot s_2k_2\cdot s_1-k_1\cdot s_1k_2\cdot s_2), \nn \\
&&B_7=s_1\cdot s_2[(p_1\cdot k_2)^2-(p_1\cdot k_1)^2]+
q\cdot s_2(p_1\cdot k_1k_1\cdot s_1-p_1\cdot k_2k_2\cdot s_1), \nn \\
&&B_8=-q\cdot s_1(k_1\cdot p_1k_2\cdot s_2-k_1\cdot s_2p_1\cdot k_2)-
s_1\cdot s_2[(p_1\cdot k_2)^2-(p_1\cdot k_1)^2]. 
\label{eq:eqp99}
\ea 
Therefore:
\ba
&&m^2B_2+q\cdot s_1B_3+q\cdot s_2B_4+B_5+B_6+B_7+B_8 =\nn \\
&&~~=2m^2(1-\tau )(k_1\cdot s_2k_2\cdot s_1-k_1\cdot s_1k_2\cdot s_2)+
q\cdot s_2(p_1\cdot k_2k_2\cdot s_1-p_1\cdot k_1k_1\cdot s_1)\nn \\ 
&&~~+q\cdot s_1(p_1\cdot k_1k_2\cdot s_2-p_1\cdot k_2k_1\cdot s_2)-
\frac{q^2}{2}q\cdot s_1(k_2-k_1)\cdot s_2. 
\label{eq:eqp8b}
\ea
So, Eq. (\ref{eq:eqp7}) for  $S^{(i)}(s_1, s_2)$ takes the form
\be
S^{(i)}(s_1, s_2)=4A_{2\gamma}\left [m^2G_E^*(k_1\cdot s_2
k_2\cdot s_1-k_1\cdot s_1k_2\cdot s_2)+\frac{1}{\tau -1}(G_M-G_E)^*
P\cdot k_1q\cdot s_1q\cdot s_2\right], 
\label{eq:eqp11}
\ee
where the following equalities were applied:
$$q\cdot s_2k_2\cdot s_1-q\cdot s_1k_1\cdot s_2=
q\cdot s_1k_2\cdot s_2-q\cdot s_2k_1\cdot s_1, $$
$$q\cdot s_1k_1\cdot s_2-q\cdot s_2k_1\cdot s_1=
k_2\cdot s_1k_1\cdot s_2-k_1\cdot s_1k_2\cdot s_2. $$
Let us calculate the following relation:
\ba
&&s_1\cdot k_2s_2\cdot k_1-s_1\cdot k_1s_2\cdot k_2=\nn \\
&&\hspace*{1true cm}(Es_{10}+{\vec k}{\vec s_1})(Es_{20}-{\vec k}{\vec s_2})-
(Es_{10}-{\vec k}{\vec s_1})(Es_{20}+{\vec k}{\vec s_2})
=2E(s_{20}{\vec k}{\vec s_1}-s_{10}{\vec k}{\vec s_1})\nn \\
&&\hspace*{1true cm}=
-\frac{q^2}{2}\frac{p}{m}\left  [\left (\frac{E}{m}\cos\theta \xi_{1z}
-\sin\theta \xi_{1x}\right)\xi_{2z}+\left (\frac{E}{m}\cos\theta \xi_{2z}
-\sin\theta \xi_{2x}\right )\xi_{1z}\right  ]. \label{eq:eqp9b}
\ea
Then the contraction Eq. (\ref{eq:eqp11}) can be expressed in terms of the
polarization unit vectors ${\vec \xi}_1$ and ${\vec \xi}_2$ as follows
\be
S^{(i)}(s_1, s_2)=-4A_{2\gamma}pq^2\left  [-\frac{m}{2}\sin\theta G_E^*(
\xi_{1x}\xi_{2z}+\xi_{2x}\xi_{1z})+E\cos\theta G_M^*\xi_{1z}\xi_{2z}
\right  ]. 
\label{eq:eqp12}
\ee
The part of the differential cross section proportional to the polarization
correlation can be written as
\be
\frac{d\sigma (s_1, s_2)}{d\Omega}=
\frac{1}{2}\frac{\alpha^2\beta}{4q^6}\left  [S^{(v)}(s_1, s_2)+
2ReS^{(i)}(s_1, s_2)\right  ], 
\label{eq:eqp13}
\ee
where the factor $1/2$ results from averaging over the positron
polarizations. The factor corresponding to the average over the electron polarization states is already
taken into account in the $L_{\mu\nu}^{(v,i)}(0)$ tensors.
Eq. (\ref{eq:eqp13}) can be written as a function of the polarization of the final hadrons as: 
\be
\frac{d\sigma (s_1, s_2)}{d\Omega}=\frac{1}{4}
\frac{d\sigma_{un}}{d\Omega}\left  [P_{xx}\xi_{1x}\xi_{2x}+
P_{yy}\xi_{1y}\xi_{2y}+P_{zz}\xi_{1z}\xi_{2z}+
P_{xz}\xi_{1x}\xi_{2z}+P_{zx}\xi_{1z}\xi_{2x}\right  ], 
\label{eq:eqp14}
\ee
where the components of the polarization correlation tensor $P_{ik}$ are
\ba
P_{xx}&=&\frac{\sin^2\theta }{\tau D}\left  [\tau (|G_M|^2+
2ReG_M\Delta G_M^*)+|G_E|^2+2ReG_E\Delta G_E^*\right  ], \nn \\ 
P_{yy}&=&\frac{\sin^2\theta }{\tau D}\left  [|G_E|^2+2ReG_E\Delta
G_E^*-\tau (|G_M|^2+2ReG_M\Delta G_M^*)\right  ],  \nn \\
P_{zz}&=&\frac{1}{\tau D}\left  [\tau (1+\cos^2\theta )(|G_M|^2+
2ReG_M\Delta G_M^*) \right . \nn \\
&&\left . -\sin^2\theta (|G_E|^2+2ReG_E\Delta G_E^*)-
4\sqrt{\tau (\tau -1)}\cos\theta ReG_MA_{2\gamma}^*\right  ],  \nn \\
P_{xz}&=&P_{zx}=-2\frac{\sin\theta }{\sqrt{\tau }D}\left  [\cos\theta
Re(G_MG_E^*+G_M\Delta G_E^*+G_E\Delta G_M^*)
 \right . \nn \\
&&\left . 
~~~~~~~~-\sqrt{\frac{\tau -1}{\tau }}ReG_EA_{2\gamma}^*\right  ]. 
\label{eq:eqi1b}
\ea
\subsection{Spin correlations: longitudinally polarized electron beam}

Let us consider now the case of  longitudinally polarized electron beam,
while the positron beam is unpolarized.

Then the contraction of the vector parts of the leptonic and hadronic
tensors can be written as
\ba 
S^{(v)}(s_1, s_2, \lambda_e)&=&L_{\mu\nu}^{(v)}(\lambda_e)
H_{\mu\nu}^{(v)}(s_1, s_2)= \nn \\
&=&
2i\lambda_e<\mu\nu qk_2>
\left  [C_1g_{\mu\nu}+C_2P_{\mu}P_{\nu}+C_3(P_{\mu}s_{1\nu}+
P_{\nu}s_{1\mu})+C_4(P_{\mu}s_{2\nu}+P_{\nu}s_{2\mu})\right .\nn \\
&&\left .+C_5(s_{1\mu}s_{2\nu}+s_{1\nu}s_{2\mu})+iC_6(P_{\mu}s_{1\nu}-
P_{\nu}s_{1\mu})+iC_7(P_{\mu}s_{2\nu}-P_{\nu}s_{2\mu})\right  ]\nn \\
&=&-4\lambda_e(C_6<Ps_1qk_2>+C_7<Ps_2qk_2>),\label{eq:eqpa}
\ea 
where we took into account the fact that the contractions of the antisymmetrical
leptonic tensor with the symmetrical parts of the hadronic tensor, which are proportional
to the structure functions $C_1-C_5$, vanish.

As an example, let us calculate the contraction $<Paqk_2>$ where $a_{\mu}$
is an arbitrary  four--vector:
\ba
<Paqk_2>&=&2E<Pa4k_2>=-2EP_i<ia4k_2>=2Ep<za4k_2>=-2Epk_{2i}<za4i>= \nn \\
&=&2Epk_{i}<za4i>=-2EpE\sin\theta <za4x>=2E^2p\sin\theta <zy4x>a_y
\nn \\
&=&
2E^2p\sin\theta a_y. 
\label{eq:eqpb}
\ea
Using this result one can easily obtain that
$$<Ps_1qk_2>=2E^2p\sin\theta \xi_{1y}, \ \
<Ps_2qk_2>=2E^2p\sin\theta \xi_{2y}. $$
Substituting the expressions for the structure functions $C_6$ and $C_7$, Eq. (\ref{eq:eqi4}):
\ba
S^{(v)}(s_1, s_2, \lambda_e)&=&-4\lambda_ep\sin\theta \frac{q^2}{2}
\frac{1}{\tau -1}ImG_MG_E^*(q\cdot s_2\xi_{1y}-q\cdot s_1\xi_{2y}) \nn\\
&=&-2\lambda_ep\sin\theta \frac{q^2}{\tau -1}ImG_MG_E^*
(-2E\frac{p}{m}\xi_{1y}\xi_{2z}-2E\frac{p}{m}\xi_{1z}\xi_{2y}) \nn\\
&=&\lambda_e\sin\theta \frac{q^4}{\sqrt{\tau}}ImG_MG_E^*
(\xi_{1y}\xi_{2z}+\xi_{1z}\xi_{2y}). 
\label{eq:eqpba}
\ea
The contraction of the interference parts of the leptonic and hadronic
tensors can be written as
\ba
S^{(i)}(s_1, s_2, \lambda_e )&=&L_{\mu\nu}^{(i)}(\lambda_e )
H_{\mu\nu}^{(i)}(s_1, s_2) \nn\\
&=&\lambda_e \left  [q^2g_{\mu\nu}-2(k_{1\mu}k_{2\nu}+k_{1\nu}k_{2\mu})\right  ]
iA_{2\gamma}\Bigl[\frac{2}{\tau -1}(G_M-G_E)^*P_{\nu}<\mu s_1s_2P> \nn\\
&& +G_M^*(m^2<\mu\nu s_1s_2>+
q\cdot s_1<\mu\nu s_2p_1>+q\cdot s_2<\mu\nu s_1p_2> \label{eq:eqp14a}\\
&& 
- p_{1\mu}<\nu s_1s_2p_2>-p_{2\nu}<\mu s_1s_2p_1>-s_{2\mu}<\nu p_2s_1p_1>-
s_{1\nu}<\mu p_2s_2p_1>)\Bigr ]. 
\nn
\ea
The contractions of the symmetrical unpolarized lepton tensor and
antisymmetrical parts of the interference hadronic tensor are equal
to zero. As a result we have
\ba
S^{(i)}(s_1, s_2, \lambda_e )&=&-i\lambda_e A_{2\gamma}\left\{ \frac{4}{\tau -1}
(G_M-G_E)^*\left  [-q^2<Ps_1s_2P>+P\cdot k_1<k_2s_1s_2P> \right .\right .
\nn\\
&&
\left .+P\cdot k_2<k_1s_1s_2P>\right ]\label{eq:eqp15}
\\
&&
+ G_M^*\left  [q^2<p_1s_1s_2p_2>-2(p_1\cdot k_1<k_2s_1s_2p_2>+
p_1\cdot k_2<k_1s_1s_2p_2>)\right .\nn\\
&&+q^2<p_2s_1s_2p_1>-2(p_2\cdot k_2<k_1s_1s_2p_1>+
p_2\cdot k_1<k_2s_1s_2p_1>) \nn\\
&&+q^2<s_2p_2s_1p_1>-2(k_1\cdot s_2<k_2p_2s_1p_1>+
k_2\cdot s_2<k_1p_2s_1p_1>) \nn\\
&&+q^2<s_1p_2s_2p_1>-2(k_2\cdot s_1<k_1p_2s_2p_1>+
k_1\cdot s_1<k_2p_2s_2p_1>)\Big ]\Big \}. \nn
\ea
The term $<Ps_1s_2P>$ vanishes, since it is the product of the antisymmetrical
tensor $\varepsilon_{\mu\nu\rho\sigma}$ and the symmetrical tensor $P_{\mu}P_{\nu}$.
To simplify this expression one can use the following property of $<abcd>$: 
any permutation of the neighboring variables gives
factor $(-1)$, for example, $<abcd>=-<acbd>$. 
Four terms, proportional to $q^2$, are canceled out. Using
the conservation of four--momenta in the reaction, $k_1+k_2=p_1+p_2,$ one
can prove that $p_1\cdot k_1=p_2\cdot k_2$ and $p_1\cdot k_2=p_2\cdot k_1$.
Taking into account these relations, we can rewrite Eq. (\ref{eq:eqp15}) in the form:
\ba
&&S^{(i)}(s_1, s_2, \lambda_e )=-i\lambda_e A_{2\gamma}
\left\{ 4\frac{P\cdot k_1}
{\tau -1}<(k_2-k_1)s_1s_2P>(G_M-G_E)^* \right .\nn \\
&&
~~~-2G_M^*\left  [p_1\cdot k_1(
<s_1s_2k_1p_1> +<s_1s_2k_2p_2>)
+p_1\cdot k_2(<s_1s_2k_1p_2>\right .
\nn\\
&& ~~~
+<s_1s_2k_2p_1>)+k_1\cdot s_1<s_2k_2p_2p_1>+k_2\cdot s_1<s_2k_1p_2p_1>\nn\\
&& ~~~
\left . 
+k_2\cdot
s_2<s_1k_1p_2p_1>+
k_1\cdot s_2<s_1k_2p_2p_1>\right  ]\Bigr \}. \label{eq:eqp16} 
\ea
One can also obtain the following relations
\ba
<s_2k_1p_2p_1>&=&<s_2(q-k_2)p_2p_1>=<s_2qp_2p_1>-<s_2k_2p_2p_1>=
-<s_2k_2p_2p_1>, \nn\\
<s_1k_2p_2p_1>&=&<s_1(q-k_1)p_2p_1>=<s_1qp_2p_1>-<s_1k_1p_2p_1>=
-<s_1k_1p_2p_1>, \nn\\
<s_1s_2k_1p_2>&=&<s_1s_2(q-k_2)p_2>=<s_1s_2qp_2>-<s_1s_2k_2p_2>, \nn\\
<s_1s_2k_2p_1>&=&<s_1s_2(q-k_1)p_1>=<s_1s_2qp_1>-<s_1s_2k_1p_1>. 
\label{eq:eqp17}
\ea
Therefore
\ba
&&k_1\cdot s_1<s_2k_2p_2p_1>+k_2\cdot s_1<s_2k_1p_2p_1>+
k_2\cdot s_2<s_1k_1p_2p_1>+k_1\cdot s_2<s_1k_2p_2p_1> \nn\\
&&~~=(k_1-k_2)\cdot s_1<s_2k_2p_2p_1>+(k_2-k_1)\cdot s_2<s_1k_1p_2p_1>,  \nn\\
&&p_1\cdot k_1(<s_1s_2k_1p_1>+<s_1s_2k_2p_2>)+
p_1\cdot k_2(<s_1s_2k_1p_2>+<s_1s_2k_2p_1>) \nn\\
&&~~=p_1\cdot k_1(<s_1s_2k_1p_1>+<s_1s_2k_2p_2>)+
p_1\cdot k_2(<s_1s_2qp_2>
\nn\\
&&~~~~~~~
-<s_1s_2k_2p_2>+<s_1s_2qp_1>- <s_1s_2k_1p_1>)
\nn\\
&&~~
=p_1\cdot (k_1-k_2)(<s_1s_2k_1p_1>+<s_1s_2k_2p_2>)+
p_1\cdot k_2<s_1s_2q(p_2+p_1)> \nn\\
&&~~=p_1\cdot (k_1-k_2)(<s_1s_2k_1p_1>+<s_1s_2k_2p_2>)+
p_1\cdot k_2<s_1s_2qq> \nn\\
&&~~=p_1\cdot (k_1-k_2)(<s_1s_2k_1p_1>+<s_1s_2k_2p_2>). 
\label{eq:eqp18}
\ea 
Eq. (\ref{eq:eqp16}) for $S^{(i)}(s_1, s_2, \lambda_e )$ can be simplified to
\ba
S^{(i)}(s_1, s_2, \lambda_e )&=&-i\lambda_e A_{2\gamma}\left\{
4\frac{P\cdot k_1}{\tau -1}<(k_2-k_1)s_1s_2P>(G_M-G_E)^* \right . \nn\\
&&-2G_M^*\left  [p_1\cdot (k_1-k_2)(<s_1s_2k_1p_1>+
<s_1s_2k_2p_2>) \right .\nn\\
&&\left .
+(k_1-k_2)\cdot s_1<s_2k_2p_2p_1>
+(k_2-k_1)\cdot s_2<s_1k_1p_2p_1>\right ]\Bigr  \}. \label{eq:eqp19}
\ea 
The following equalities hold:
\ba
&&<s_1k_1p_2p_1>=<s_1k_1(q-p_1)p_1>=<s_1k_1qp_1>=<qs_1k_1p_1>, \nn\\
&&
<s_2k_2p_2p_1>=<s_2k_2(q-p_1)p_1>=<s_2k_2qp_1>=<s_2(q-k_1)qp_1>=
-<qs_2k_1p_1>.\nn
\ea 
It is convenient to calculate some auxiliary terms
\ba
<qak_1p_1>&=&2E<4ak_1p_1>=-2E<4ak_1i>p_i=-2Ep<4ak_1z> \nn \\
&=&2Ep<4aiz>k_i=-2EpE \sin\theta <4axz>\nn \\
&=&2E^2p\sin\theta <4yxz>a_y=
2E^2p\sin\theta a_y,   \nn\\
<s_1s_2Pa>&=& a_0p(s_{1x}s_{2y}-s_{1y}s_{2x})-pa_x(s_{10}s_{2y}-
s_{1y}s_{20}). \label{eq:eqp20}
\ea
The last relation is obtained assuming that $a_y=0$.  
Other useful relations are:
\ba
&&<s_1s_2k_1p_1>+<s_1s_2k_2p_2>=<s_1s_2k_1p_1>+
<s_1s_2(q-k_1)p_2>\nn\\
&&~~~~=<s_1s_2qp_2> +<s_1s_2k_1(p_1-p_2)>=<s_1s_2qp_2>+2<s_1s_2Pk_1>,\nn\\
&&<s_1s_2qp_2>=2E<s_1s_24p_2>=2Ep<s_1s_24z>=
2Ep(s_{1y}s_{2x}-s_{1x}s_{2y}), \nn\\
&&<s_1s_2Pk_1>=Ep(s_{1x}s_{2y}-s_{1y}s_{2x})+
pE\sin\theta (s_{10}s_{2y}-s_{1y}s_{20}),  \nn\\
&&<s_1s_2Pk_2>=Ep(s_{1x}s_{2y}-s_{1y}s_{2x})-
pE\sin\theta (s_{10}s_{2y}-s_{1y}s_{20}). \label{eq:eqp21}
\ea
Inserting (\ref{eq:eqp20},\ref{eq:eqp21}) in (\ref{eq:eqp16}), one finds:
\ba
&&S^{(i)}(s_1, s_2, \lambda_e )=-i\lambda_e A_{2\gamma}\Bigl \{
2mpq^2\sin\theta \cos\theta (G_M-G_E)^*(\xi_{1y}\xi_{2z}+
\xi_{1z}\xi_{2y}) \nn\\
&&~~-2G_M^*\left  [Epq^2\sin\theta \left  (\frac{E}{m}\cos\theta \xi_{2z}
-\sin\theta \xi_{2x} \right )\xi_{1y}+Epq^2\sin\theta
\left(\frac{E}{m}\cos\theta \xi_{1z}-\sin\theta \xi_{1x}\right  )\xi_{2y} \right  .\nn\\
&&~~~~~~~~~~~~~~\left .\left . -p^2q^2\frac{p}{m}\sin\theta \cos\theta
\left(\xi_{1y}\xi_{2z}+\xi_{1z}\xi_{2y}\right )\right  ]\right \} \nn\\
&&=-2i\lambda_e pq^2\sin\theta A_{2\gamma}\left  [E\sin\theta G_M^*
(\xi_{1x}\xi_{2y}+\xi_{1y}\xi_{2x})-m\cos\theta G_E^*
(\xi_{1y}\xi_{2z}+\xi_{1z}\xi_{2y})\right  ]. \label{eq:eqp21a}
\ea
Then the part of the differential cross section, proportional to the
polarization correlation and to the longitudinal polarization of the
electron beam, can be written as
\be
\frac{d\sigma (s_1, s_2, \lambda_e )}{d\Omega}=
\frac{1}{2}\frac{\alpha^2\beta}{4q^6}\left  [S^{(v)}(s_1, s_2, \lambda_e )+
2ReS^{(i)}(s_1, s_2, \lambda_e )\right  ], \label{eq:eqp21b}
\ee
where extra factor $1/2$ comes from averaging only over the positron
polarizations since the electron beam is polarized.

The correlation polarization tensor, for the case of the
longitudinally polarized electron beam, is defined as follows
\be\frac{d\sigma (s_1, s_2, \lambda_e )}{d\Omega}=\frac{\lambda_e}{2}
\frac{d\sigma_{un}}{d\Omega}\left  [P_{xy}\xi_{1x}\xi_{2y}+
P_{yx}\xi_{1y}\xi_{2x}+P_{yz}\xi_{1y}\xi_{2z}+
P_{zy}\xi_{1z}\xi_{2y}\right  ],\label{eq:eqp22}
\ee 
where the components of the polarization correlation tensor $P_{ik}$ are
\ba
P_{xy}&=P_{yx}=&-\frac{1}{D}\sqrt{\frac{\tau -1}{\tau}}\sin^2\theta
ImG_MA_{2\gamma}^*, \nn \\
P_{zy}&=P_{yz}=&\frac{\sin\theta }{\sqrt{\tau }D}Im(G_MG_E^*+
G_M\Delta G_E^*-G_E\Delta G_M^*+ \nn \\
&&+\sqrt{\frac{\tau -1}{\tau}}\cos\theta G_EA_{2\gamma}^*), \label{eq:eqp23}
\ea 
and we used the relation $Re(iA)=Im A^*.$

One can easily verify that the following relation holds:
$$P_{xx}+P_{yy}+P_{zz}=1. $$
The components of the tensor describing the polarization correlations
$P_{xx}, $ $P_{yy}, $ $P_{zz},$ $P_{xz},$ and $P_{zx}$ are T--even
observables, whereas the components $P_{xy}, $ $P_{yx}, $ $P_{yz},$ and
$P_{zy}$ are T--odd ones.

In the Born approximation the expressions for the T--odd polarization
correlations coincide with the corresponding components of the
polarization correlation tensor of baryon $B$ and antibaryon $\bar B$
created by the one--photon--exchange mechanism in the $e^+e^-\to B\bar B$
process: Let us write explicitly these expressions: 
\begin{eqnarray*}
P_{xx}&=&\frac{\sin^2\theta }{\tau D} [\tau |G_M|^2+|G_E|^2
], \nonumber \\
P_{yy}&=&\frac{\sin^2\theta }{\tau D}
 [|G_E|^2-\tau |G_M|^2], \nonumber \\
P_{zz}&=&\frac{1}{\tau D }
[\tau (1+\cos^2\theta )|G_M|^2-\sin^2\theta |G_E|^2],  \nonumber \\
P_{xz}&=&P_{zx}=-\frac{\sin 2\theta }{\sqrt{\tau }D} Re[G_MG_E^*],
\nonumber \\
P_{xy}&=&P_{yx}=0,\nonumber \\
P_{zy}&=&P_{yz}=\frac{\sin\theta }{\sqrt{\tau }D}Im(G_MG_E^*).\nonumber 
\end{eqnarray*}

\subsubsection{$2\gamma$ mechanism and double spin polarization coefficients}

The relative contribution of the interference terms (between one- and
two--photon--exchange terms) in these observables will increase as
the value $q^2$ becomes larger since it is expected that the TPE amplitudes
decrease more slowly with $q^2$ compared with the nucleon FFs.

At the reaction threshold, the polarization correlation tensor components have
some specific properties:

- All correlation coefficients (both T--odd and T--even) do not depend on
the function $A_{2\gamma}$.

- In the Born approximation the $P_{yy}$ observable is zero, but the presence 
of the TPE term leads to a non--zero value, determined by the quantity
$2Re(G_E\Delta G_E^*-G_M\Delta G_M^*)$.

- At the scattering angle $\theta = 90^0$  the relation $P_{yy}+
P_{zz}=0$ holds.

- The $P_{xy}$ and $P_{yx}$ observables are zero, and $P_{yz}$ and $P_{zy}$
observables are determined by the TPE term only, namely by the quantity
$ImG_{M}(\Delta G_E-\Delta G_M)^*.$

\section{Conclusions}

A detailed model independent derivation of experimental observables in $e^++e^-\to N+\bar N$ is given here with a specific pedagogical aim toward  students at PhD level. 

The present analysis is also a guideline for the experimental 
investigation of the TL nucleon FFs planned in near future at Laboratories 
where electron and positron beams are available.

The short bibliography given below contains more discussion and results, based on similar 
ideas and formalism can be found, for this reaction and all crossed reactions 
as well. 
A wide literature exist on this subject, since the first studies, in the 
fifties.
The reader is invited to consult references contained in the quoted paper,  to access other related works, experimental or theoretical. 

This formalism has been extended to spin one hadrons.

\section{Aknowledgments}
The model independent derivation of the properties of two photon exchange in different electron hadron reactions was started by a collaboration with Prof. M. P. Rekalo. 
The work of one of us (G.I.G) is supported in part by the INTAS grant, under Ref. Nr. 05-1000008-8328.
The Slovak Grant Agency for Sciences VEGA is acknowledged by C.A., for support under Grant N. 2/4099/26.

\end{document}